\newcommand{\be}{\begin{equation}}\newcommand{\ee}{\end{equation}}
\newcommand{\bea}{\begin{eqnarray}}\newcommand{\eea}{\end{eqnarray}}

\newcommand{\nn}{\nonumber}\newcommand{\p}[1]{(\ref{#1})}
\newcommand{\lpr}{{\cal L}^{+4}}

\newcommand{\fracds}[2]{\displaystyle \frac{#1}{#2} }
\documentstyle[12pt]{article}

\begin{document}

\thispagestyle{empty}
\begin{flushright}   JHU-TIPAC-920023\\
ENSLAPP-L-405-92 \\
MPI-Ph/92-85\\ October 1992\\ hep-th/9212155
\end{flushright}

\bigskip\bigskip\begin{center} {\bf
\Huge{Harmonic space and quaternionic manifolds}}
\end{center}  \vskip 1.0truecm

\centerline{\bf A. Galperin${}^{(a)\dag}$, E. Ivanov${}^{(b)\dag}$
and O. Ogievetsky${}^{(c)\ddag}$}
\vskip5mm
\centerline{${}^{(a)}$ Department of Physics and Astronomy}
\centerline{Johns Hopkins University, Baltimore, MD 21218}
\vskip5mm
\centerline{${}^{(b)}$Laboratoire de Physique Theorique ENSLAPP, ENS Lyon}
\centerline{46 Allee d'Italie, F-69364 Lyon Cedex 07, France}
\vskip5mm
\centerline{${}^{(c)}$Max-Planck-Institut f\"{u}r
 Physik, Werner-Heisenberg-Institut}
\centerline{F\"ohringer Ring 6, 8000 Munich 40, Germany}

\bigskip \nopagebreak \begin{abstract}
\noindent We find a principle of harmonic analyticity underlying the
quaternionic
(quaternion-K\"ahler) geometry and solve the differential constraints which
define this geometry. To this end the original $4n$-dimensional
quaternionic manifold is extended to a bi-harmonic space. The latter
includes additional harmonic coordinates associated with both the tangent
local $Sp(1)$ group and an extra rigid $SU(2)$
group rotating the complex structures. Then the constraints can be
rewritten as integrability conditions for the existence of an analytic
subspace in the bi-harmonic space and solved in terms  of two unconstrained
potentials on the
analytic subspace. Geometrically, the potentials have the meaning of
vielbeins associated with the harmonic coordinates.
We also establish  a one-to-one correspondence between the quaternionic
spaces and off-shell $N=2$ supersymmetric sigma-models coupled to $N=2$
supergravity. The general $N=2$ sigma-model Lagrangian when written in the
harmonic superspace is composed of the quaternionic potentials.
Coordinates of the analytic subspace are identified with
superfields describing $N=2$ matter hypermultiplets and a compensating
hypermultiplet of $N=2$ supergravity.
As an illustration we present the potentials for the
symmetric quaternionic spaces.

\end{abstract}

\bigskip \nopagebreak \begin{flushleft} \rule{2
in}{0.03cm} \\ {\footnotesize \ ${}^{\dag}$  On leave from the Laboratory
of Theoretical Physics, JINR, Dubna, Head Post Office, P.O. Box 79,
101000 Moscow, Russia}
\\ {\footnotesize \ ${}^{\ddag}$ On leave from P.N. Lebedev Physical
 Institute, Theoretical Department, 117924 Moscow, Leninsky
 prospect 53, Russia}
\end{flushleft}

\newpage\setcounter{page}1

\section{Introduction}

In the present paper we complete a study of gauge theories with
self-duality type constraints  within the framework of harmonic space
approach. This study was initiated in papers  \cite{{a1},{a2}} where
the self-dual Yang-Mills fields and hyper-K\"ahler (HK) metrics
(generalizing to $4n$ dimensions the
self-dual Einstein equations) were considered.
There the concept of harmonic analyticity,
originally introduced in the framework of $N=2$ supersymmetry \cite{a3},
was applied to purely bosonic self-dual Yang-Mills and Einstein
theories in the four-dimensional space $R^{4}$ and their generalizations
to $R^{4n}$.

The harmonic superspace
approach \cite{a3} is based on an extension of
the original real $N=2$ superspace by harmonics
$u^{\pm}_{i}, \;\; u^{+i}u^-_i=1, \;\; i=1,2$ parametrizing
the sphere $S^2=SU(2)/U(1)$ (where $SU(2)$ is the automorphism group of
$N=2$ supersymmetry). The reasons for this are:
({\it i})  the extended superspace has
an analytic subspace of smaller Grassmann dimension
(which essentially involves harmonics);
({\it ii}) the main constraints of all $N=2$
supersymmetric theories can be interpreted as
integrability conditions for this
harmonic analyticity; ({\it iii}) these constraints can be solved in
terms  of some arbitrary analytic potentials.

Similarly, in self-dual Yang-Mills  and Einstein theories one
({\it i}) extends the original real $R^{4n}$ space (whose tangent group
is $SU(2) \times Sp(n)$ and holonomy group is contained in $Sp(n)$)
by the sphere $S^2=SU(2)/U(1)$
coordinatized by the harmonics $u^\pm_i$; ({\it ii}) interprets
the original
equations as integrability conditions for the existence
of an analytic subspace which essentially involves harmonics and contains a
half of the original $R^{4n}$ coordinates; ({\it iii}) solves these
equations in terms of some arbitrary analytic potentials \cite{{a1},{a2}}.
Thus, analyticity means a half of anticommuting coordinates in supersymmetry
and a half of commuting coordinates in self-dual theories.
We emphasize that the harmonic analytic potentials encode the full
information about the original (constrained) theory
and provide a tool of classifying and computing HK metrics and
self-dual Yang-Mills fields.

An interesting application of HK manifolds in
field theory is connected with their role as the target spaces of $N=2$
supersymmetric sigma models \cite{a5}. The unconstrained formulation of
these manifolds given in \cite{a2} makes this connection transparent.
It turns out that the general superfield Lagrangian of $N=2$ sigma models
derived in \cite{{a6},{a7}} using the harmonic superspace approach,
is in one-to-one correspondence with the
HK potentials. The basic $N=2$ matter superfields
(analytic $Q^{+}$ hypermultiplets) together with
harmonics $u^\pm_i$ are recognized as
coordinates of the abovementioned analytic subspace
of the extended HK target manifold $R^{4n}\times S^{2}$.
Both the target harmonic space and the harmonic superspace (on which
the hypermultiplets $Q^{+}$ are defined) contain the {\it same}
set of harmonics $u^{\pm}_{i}$.
This implies identification of the automorphism
group $SU(2)$ of $N=2$ supersymmetry  with the rigid $SU(2)$ group
rotating three complex structure of the target HK manifold.

Thus, the harmonic space formulation of HK geometry on the one
hand, and the harmonic superspace description of $N=2$ matter on the other,
make the relation between this geometry and $N=2$ supersymmetry as
clear as that between the K\"ahler geometry and $N=1$ supersymmetry
\cite{a8}. Moreover, the fact that the $N=2$ matter superfield Lagrangian
of ref. \cite{{a6},{a7}} is composed of HK potentials,
provides a direct proof \cite{a2} that this Lagrangian describes the most
general matter self-couplings in the rigid $N=2$ supersymmetry.

When $N=2$ sigma model is put in a supergravity background, its bosonic target
space is known to be quaternionic \cite{a9} rather than hyper-K\"ahler as in
the flat case. The most general off-shell action of such sigma models
was constructed in \cite{a10} in the framework of harmonic superspace.
By analogy with the rigid case, the relevant superfield Lagrangian was
conjectured to be composed of unconstrained
potentials underlying the quaternionic geometry.
As was observed
in \cite{{a10},{a11}}, this object reveals a number of new features
compared to its rigid $N=2$ supersymmetric prototype. For instance,
the harmonics which
appear explicitly in the Lagrangian and so are expected to be related to
the target space geometry do not coincide with the harmonics present in the
harmonic superspace. Instead they are expressed through them and the
compensating analytic superfield $q_i^{+}$ of $N=2$ supergravity.

In order to understand the structure and symmetries of the superfield
$N=2$ sigma model action of ref. \cite{a10} and to prove its
conjectured status as the most general action of matter in
local $N=2$ supersymmetry, one clearly needs an unconstrained
geometric formulation of the quaternionic geometry similar to
that of the HK case \cite{a2}. Such a formulation is
given in the present paper.

Quaternionic manifolds (see, e.g. \cite{{a13},{a16},{a12},{a14}}) are
$4n$ dimensional Riemannian manifolds with
the holonomy group contained in $Sp(1)\times Sp(n)$. Locally, the
quaternionic manifolds (like the HK manifolds )
have three almost complex structures that obey
the well-known Clifford algebra relation. However, now  these structures are
not covariantly constant, the
$Sp(1)$ part of the holonomy group rotates them.
So, a distinction from
the HK case is that the complex structures on quaternionic manifolds
are in principle not defined globally and
cannot be chosen constant. Actually,
the HK manifolds can be regarded as a degenerate class of the
quaternionic ones whose holonomy group does not contain the $Sp(1)$
factor. Below we shall consider the general case with the $Sp(1)$ factor
being nontrivial. As was shown in \cite{a9}, this is the
geometry inherent to the bosonic manifolds of $N=2$ sigma models in
the supergravity background.

We will show that the quaternionic geometry, like the
HK one, is determined by a harmonic analyticity
preservation principle which naturally leads to unconstrained
analytic potentials as the fundamental objects of this geometry.
An important feature of the quaternionic
case is that an interpretation of the basic constraints as
integrability conditions for the preservation of harmonic analyticity
becomes possible only after extending the original $4n$ dimensional
manifold by two sets of harmonic coordinates.
One of them consists of the ordinary harmonics of
an extra {\it rigid} $SU(2)$ acting on the compex structures
in the same way as in the HK case. The other is new; it parametrizes the
{\it local} $Sp(1)$ group acting in the tangent space of the manifold.
In the limiting case of HK
manifolds ( i.e., when the $Sp(1)$ curvature of quaternionic manifolds is
vanishing) the new coordinates coincide with
central-charge coordinates introduced in \cite{a2} to solve the HK
constraints in an algorithmic way. Thus the geometric meaning of the
central-charge coordinates is naturally revealed in the framework of the
harmonic space formulation of the quaternionic geometry.
New harmonic variables will be shown to be
directly related to the analytic compensating hypermultiplet of $N=2$
supergravity which thus acquires a nice geometric
interpretation.

The paper is planned as follows. We begin in Sect. 2 with giving the basic
notions of the quaternionic geometry in real $4n$ dimensional space
(or $\tau$ world). We write
the defining constraints of this geometry in a form similar to the
self-dual Yang-Mills and HK geometry constraints as they have been given
in \cite{{a1},{a2}}. In Sect. 3 we introduce the concept of bi-harmonic
space and give the necessary
technical details. In Sect. 4 we interpret the quaternionic geometry
constraints as integrability conditions for the existence of a kind
of harmonic analyticity in a bi-harmonic extension of the original
manifold. Sect. 5 describes the procedure of passing to a new
representation of these constraints where the underlying analyticity
becomes manifest (so called $\lambda$-world representation).
In Sect. 6 and 7 we study the consequences of the
quaternionic geometry constraints in the $\lambda$-world. In the process
of this study in Sect. 7 we naturally come to unconstrained analytic
quaternionic potentials ${\cal L}^{+4}$ and ${\cal L}^{+}_{\mu}$ which
emerge as harmonic vielbeins in some $\lambda$-world covariant derivatives.
We prove that all geometric
quantities relevant to the quaternionic geometry are expressible,
like in the HK case \cite{a2},
in terms of these two basic objects (in fact only one of
them, $\lpr$, is essential; the second one, as in the
HK case, is locally a pure gauge). In Sect. 8 we treat the simplest example of
``maximally flat'' $4n$ dimensional quaternionic manifold, the homogeneous
space $Sp(n+1)/Sp(1)\times Sp(n)$. The limiting procedure allowing to
reproduce the harmonic space formulation of HK manifolds of ref. \cite{a2}
by a contraction of that for
quaternionic manifolds is described in Sect. 9.
In Sect. 10 we analyze the most general harmonic superspace off-shell action
of $N=2$
matter in the supergravity background \cite{a10} from the point of view of
the harmonic space approach to quaternionic geometry. We show that all
objects entering the action (including the supergravity compensators)
as well as the relevant equations of motion
have a clear interpretation in the framework of this approach. Thus the
latter visualizes the one-to-one correspondence between local $N=2$
supersymmetry and quaternionic geometry \cite{a9}. Finally, in Sect. 11 we
give the explicit form of the basic analytic potential $\lpr$ for all symmetric
quaternionic spaces classified by Wolf \cite{a13}.

Before turning to the presentation, it is useful to point out that in this
paper, similarly to \cite{{a1},{a2}}, we mainly use
the standard language of differential geometry
which is widely used in superfield gauge theories and operates with such
notions as vielbeins, curvatures, torsions, tangent space, etc. All
concepts and results given below are certainly transferrable to the
language of complex geometry and twistor theory (the latter has been
used to solve self-dual Einstein and Yang-Mills equations in the pioneering
papers \cite{Pen}, \cite{Ward} and
to study quaternionic manifolds in \cite{a12}).
One of our aims here is to demonstrate that
the harmonic
analyticity which has already proved its efficiency
in a wide range of problems, both in supersymmetric and purely bosonic
theories, is appropriate for the quaternionic geometry as well.
In our presentation we closely follow the logic and motivation
of papers \cite{{a1},{a2}} so the reader may consult them for simpler
examples.

\section{Quaternionic geometry in the real basis}

We start with a $4n$ dimensional Riemann manifold $M$ with local coordinates
$\{ x^{\mu k}\}, \;\;
\mu=1,2,\ldots 2n; \;\;k=1,2$. One of the definitions of
quaternionic geometry \cite{{a9},{a12},{a13},{a14}} restricts the holonomy
group to a subgroup of $Sp(1)\times Sp(n)$. Hence we can choose from the
very beginning the tangent group to be $Sp(1)\times Sp(n)$. So,
the tensor fields defined on the manifold $\{ x^{\mu k}\}$
undergo gauge transformations both in their $Sp(n)$ and $Sp(1)$
indices
\bea
\delta\phi_{\alpha\beta\ldots ij\ldots}(x) &\equiv&
 \phi'_{\alpha\beta\ldots ij\ldots}(x+\delta x)-
\phi_{\alpha\beta\ldots ij\ldots}(x) \nn \\
 & =&
\tau_\alpha{}^{\alpha'}(x)\phi_{\alpha'\beta\ldots ij\ldots}(x)+
\tau_\beta{}^{\beta'}(x)\phi_{\alpha\beta'\ldots ij\ldots}(x)+\ldots \nn
\\  & +&
\tau_i{}^{i'}(x)\phi_{\alpha\beta\ldots i'j\ldots}(x)+
\tau_j{}^{j'}(x)\phi_{\alpha\beta\ldots ij'\ldots}(x)+\ldots
\nn \\
\delta x^{\mu i} &=& \tau^{\mu i}(x)\;. \label{5.2.1}
\eea
Correspondingly, the covariant derivative is given by
\bea
{\cal D}_{\alpha i} = e^{\mu k}_{\alpha i}\partial_{\mu k}-
\omega_{\alpha i(\sigma\rho)}\Gamma^{(\sigma\rho)}-
\omega_{\alpha i(lk)}\Gamma^{(lk)} \equiv  \nabla_{\alpha i}-
 \omega_{\alpha i(\sigma\rho)}\Gamma^{(\sigma\rho)}-
\omega_{\alpha i(lk)}\Gamma^{(lk)}\;. \label{5.3.1}
\eea
Here $\omega_{\alpha i(\sigma\rho)}$ and
$\omega_{\alpha i(lk)}$ are the $Sp(n)$ and $Sp(1)$ connections,
$\Gamma^{(\sigma\rho)}$ and $\Gamma^{(lk)}$ are $Sp(n)$ and $Sp(1)$
generators
\bea
[\Gamma^{(\sigma\rho)},\Gamma^{(\delta\kappa)}]_A{}^C =
{1\over 2}( \Omega^{\sigma\delta}\Gamma^{(\rho\kappa)}+
\Omega^{\sigma\kappa}\Gamma^{(\rho\delta)}  + \Omega^{\rho\delta}
\Gamma^{(\sigma\kappa)}+
\Omega^{\rho\kappa}\Gamma^{(\sigma\delta)})_A{}^C
\eea
(the similar relation holds for $\Gamma^{(lk)}$, with the $Sp(n)$
invariant tensor $\Omega^{\rho\delta}$ replaced by $\epsilon^{lk}$).
As a rule we  deal with the fundamental spinor representations
of $Sp(n)$ and $Sp(1)$
\bea
({\cal D}_{\alpha i})_{\beta n}{}^{\beta' n'} =
\delta_\beta^{\beta'}\delta_n^{n'}\nabla_{\alpha i} +
\delta_n^{n'}\omega_{\alpha i\;\beta}{}^{\beta'} +
\delta_\beta^{\beta'}\omega_{\alpha i\;n}{}^{n'}\;. \label{5.3.2}
\eea
The commutator of two covariant derivatives produces
the $Sp(n)$ and $Sp(1)$ components of the curvature tensor
\bea
[{\cal D}_{\alpha i},{\cal D}_{\beta j}]_{\rho n}{}^{\rho' n'} =
\delta_n^{n'}
R_{\alpha i\;\beta j\; \rho}{}^{\rho'}+
\delta_\rho^{\rho'}R_{\alpha i\;\beta j\;n}{}^{n'}
\equiv R_{\alpha i,\;\beta j\; \rho n}^{\rho' n'}\;. \label{5.4.1}
\eea

Now, the requirement that the holonomy group of this $4n$ dimensional
Riemannian manifold (i.e. the group generated by the Riemann tensor)
belongs to  $Sp(n) \times Sp(1)$  is equivalent to the following
covariant constraints
\bea
R_{\alpha i\;\beta j\; \rho}{}^{\rho'} &=& \epsilon_{ij}
R_{\alpha\beta ; \rho}{}^{\rho'} \label{5.4.2a}\\
R_{\alpha i\;\beta j\;n}{}^{n'} &=&
\Omega_{\alpha\beta}R_{i j\;n}{}^{n'}\;. \label{5.4.2b}
\eea
The difference from the analogous
constraint defining the HK manifolds \cite{a2} lies in eq.
\p{5.4.2b}. Its r.h.s. is vanishing in the HK case, while
in the quaternionic case it corresponds to the nonvanishing $Sp(1)$
part of the holonomy group.

The Bianchi identities combined with \p{5.4.2a} and \p{5.4.2b} imply
\bea
R_{(\alpha\beta)\;(\rho\beta')} &=&R_{(\alpha\beta\rho\beta')}+
(\Omega_{\beta\rho}\Omega_{\alpha\beta'}+
\Omega_{\alpha\rho}\Omega_{\beta\beta'})R \label{5.4.3a} \\
R_{(ij)(kl)}&=&(\epsilon_{ik}\epsilon_{jl}+
\epsilon_{il}\epsilon_{jk})R\;, \;\;\;\ R={\rm constant}\;,\label{5.4.3b}
\eea
where
\bea
R\equiv {1\over 6}R_{(ij)}{}^{(ij)}\;, \;\;\;
R_{\alpha i\;\beta j}{}^{\alpha i\;\beta j}=8n(n+2)R\;. \label{5.4.4}
\eea
The constant $R$ can be positive or negative, the constraints (2.6), (2.7)
do not fix its sign. It is easy to see from eqns (2.5) - (2.9)
that quaternionic manifolds are Einstein manifolds (with cosmological
constant proportional to $R$). Hence
{\it homogeneous} quaternionic manifolds are compact in the case $R > 0$
and noncompact if $R < 0$ \cite{besse}. The scalar $Sp(1)$
curvature is by definition positive and is given by $|R_{ij}{}^{ij}| = 6 |R|$.

Note that for $n=1$ (four-dimensional manifolds) the relations
\p{5.4.2a} and \p{5.4.2b} are satisfied trivially and place no
restrictions on the geometry (correspondingly, the Bianchi identities
do not require $R_{(ijkl)}$ to vanish). In this case one is led to
impose the condition
\bea
R_{(ijkl)}=0 \label{}
\eea
"by hand". This gives rise to the generic form \p{5.4.3b} for
$R_{(ij)(kl)}$ and provides the definition of the quaternionic
geometry for $n=1$.

  Thus, irrespective of the value of $n$, the basic constraint
defining the quaternionic geometry can be written as follows
\bea
[{\cal D}_{\alpha i},{\cal D}_{\beta j}]_{\rho n}{}^{\rho' n'} =
-2\;\delta_\rho^{\rho'}\Omega_{\alpha\beta}\;R\;\Gamma_{(ij)\;n}{}^{n'}
+ \delta_n^{n'}\epsilon_{ij}[ R_{(\alpha\beta\rho}{}^{\rho')}-
R\;(\Omega_{\beta\rho}\delta_\alpha^{\rho'}+
\Omega_{\alpha\rho}\delta_\beta^{\rho'})] \label{5.5.1}
\eea
with
\be
\Gamma_{(ij)\;n}{}^{n'}={1\over 2}(\epsilon_{in}\delta_j^{n'}+
\epsilon_{jn}\delta_i^{n'})\;. \nn
\ee
Symmetrizing \p{5.5.1} with respect to $i, j$ one obtains an
equivalent equation
\be
[{\cal D}_{\alpha (i},{\cal D}_{\beta j)}]_{\rho n}{}^{\rho' n'}=
-2\;\delta_\rho^{\rho'}\Omega_{\alpha\beta}R\Gamma_{(ij)\;n}{}^{n'}\;.
\label{5.5.2}
\ee
Similar equations in the self-dual Yang-Mills and HK geometries contain
the vanishing r.h.s. \cite{{a1},{a2}}. In fact, the HK
manifolds are related to the quaternionic ones via contraction
$R\rightarrow 0$ which means the vanishing of the $Sp(1)$ curvature
$ \sim |R|$ (see also Sect.9).

We would like to understand eq. \p{5.5.2} as an integrability
condition. However, the standard trick \cite{{a1},{a2}}
with multiplying the l.h.s. by harmonics $u^{+i}u^{+j}$
does not immediately lead to this interpretation.
The reason is the presence of the nonvanishing scalar $Sp(1)$ curvature
$\sim |R|\neq 0$.

Our further strategy is as follows. First, we shall introduce the
coordinates of tangent space $Sp(1)$ group, thus representing the $Sp(1)$
part of the tangent space group by differential operators acting on these
new variables. Then the $Sp(1)$ curvature term in \p{5.5.1}, \p{5.5.2} will be
reinterpreted as a component of the {\it torsion} associated with the new
$Sp(1)$ coordinates. After this, we shall introduce the ordinary
harmonics corresponding to some extra {\it rigid} automorphism $SU(2)$
group rotating the $Sp(1)$ indices. Then it will be shown that eq. \p{5.5.2}
admits the interpretation of the integrability condition for the existence of
an appropriate (covariantly) analytic subspace in such a ``bi-harmonic''
extension of the original $4n$ dimensional manifold. Finally, we shall
solve the constraints and find the
unconstrained analytic potentials of the quaternionic geometry.

\setcounter{equation}{0}
\section{Harmonic extension of the sphere $S^2\sim Sp(1)/U(1)$}

Here we give the basic technical details on harmonic
coordinates on the  sphere $S^2\sim Sp(1)/U(1)$ and extension of the
latter by extra harmonics $u^{\pm i}$.

We parametrize  $Sp(1)$ by the harmonics $v_k{}^a$, i.e. by the $Sp(1)$
matrix in the doublet representaion:
\bea
v_k{}^av_a{}^l = -\delta_k^l\;, \;\;\; v_k{}^av_b{}^k=-\delta_b^a\;,
\;\;\;\delta v_k{}^a = \tau_k{}^l(x)v_l{}^a\;. \label{5.7.1}
\eea
The indices $i,j,k,\ldots$ correspond to the local $Sp(1)$ group (cf.
\p{5.2.1}), while the indices $a,b,\ldots$ are inert under
this group. The $Sp(1)$ generators are given by
\bea
\Gamma^{(ij)}={1\over 2}( v_a{}^i\epsilon^{jk}
{\partial \over {\partial v_a{}^k}}+
v_a{}^j\epsilon^{ik}
{\partial \over {\partial v_a{}^k}})\;, \label{5.7.2}
\eea
while the operators of the covariant differentiation on the $Sp(1)$
group manifold by
\bea
Z^{(ab)}={1\over 2}( v_k{}^a\epsilon^{bc}
{\partial \over {\partial v_k{}^c}}+
v_k{}^b\epsilon^{ac}
{\partial \over {\partial v_k{}^c}}) = -v_k{}^av_l{}^b \Gamma^{(kl)}
\label{5.7.3}
\eea
Both sets of the differential  operators
\p{5.7.2} and \p{5.7.3} form $Sp(1)\sim SU(2)$ algebras and
commute with each other, e.g
\be
[Z^{(ab)}, Z^{(cd)}]= \epsilon^{bd}Z^{(ac)}+\epsilon^{bc}Z^{(ad)}
+(a\leftrightarrow b) \label{5.7.4}
\ee
\be
[Z^{(ab)},\Gamma^{(ij)}]=0 \;.\nn
\ee
Next, let us assume that on the right index $a$ of $v_k^a$ some extra
{\it rigid} $SU(2)$ is realised
(so $v_k^a$ is a kind of ``bridge''
relating the local tangent $Sp(1)$ and this rigid $SU(2)$)
\bea
\delta v_k^a=-(\tau_r)_b^a v_k^b,
\;\;\;\;\;  (\tau_r)_b^a={\rm const} \nn
\eea
and consider a ``harmonic extension'' of the space $\{x,\; v_k^a\}$
by the harmonics $u^{\pm a}$ defined on $SU(2)/U(1)$:
\bea
\{x,\; v_k^a\} \; \Rightarrow \; \{x,\; v_k^a, \;u^{\pm a}\}, \;\;\;
u^{+a}u^-_a  =  1, \;\;\overline{u^-_a}=u^{+a}\;, \;\;
\delta u^{\pm a} = -(\tau_r)_b^a u^{\pm b}\;.\label{5.7.5}
\eea

Further we shall proceed in a close parallel with singling out
analytic subspaces in the harmonic spaces and superspaces
\cite{{a1},{a2},{a3},{a4}}.

First of all we pass to
another basis of the bi-harmonic space \p{5.7.5}
\bea
\{x,\; v_k^a, \;u^{\pm a}\} \; \Rightarrow \; \{x,\; v_k^{\pm}, \;u^{\pm
a}\}\;,
\;\;\; v_k^{\pm} =  v_k^au^{\pm a}, \;\; v^{+k}v^-_k = 1
\label{5.8.1}
\eea
\bea
Z^{\;++}&=&-{1\over 2}Z^{(ab)}u^+_au^+_b \;=\;
 v^+_k {\partial \over {\partial v_k^-}} \nn \\
Z^{\;--}&=&{1\over 2}Z^{(ab)}u^-_au^-_b \;=\;
 v^-_k {\partial \over {\partial v_k^+}} \nn \\
Z^{\;0}&=&Z^{(ab)}u^+_au^-_b \;=\;
 v^+_k {\partial \over {\partial v_k^+}} -
 v^-_k {\partial \over {\partial v_k^-}}\;. \label{5.8.2}
\eea
The commutators between the $v$-derivatives $Z$
\p{5.8.2} and the non-vanishing commutators of the latter with the harmonic
derivatives  $\partial^{\;++}=u^{+a}\partial/\partial u^{-a},
\partial^{\;--}=u^{-a}\partial/\partial u^{+a}$ and $\partial^{\;0}=
u^{+a}\partial/\partial u^{+a}-u^{-a}\partial/\partial u^{-a}$ are given
by
\bea
[Z^{\;++}, Z^{\;--}]=Z^{\;0}\;, \; \; \;
[Z^{\;0}, Z^{\;\pm\pm}]=\pm\; 2Z^{\;\pm\pm}
\label{5.8.3}
\eea
\bea
[\partial^{\;++}, Z^{\;--}]=Z^{\;0}\;, \ \ \ [\partial^{++}, Z^0]=-\;2Z^{\;0}
\nn
\eea
\bea
[\partial^{\;--}, Z^{\;++}]=-\;Z^{\;0}\;, \ \ \
[\partial^{\;--},Z^{\;0}]=2\;Z^{\;--} \nn
\eea
\bea
[\partial^{\;0}, Z^{\;\pm\pm}]=\pm \;2Z^{\;\pm\pm}\;. \label{5.8.a}
\eea

One more change of variables leads us to the following ``analytic''
basis in the space $\{x,\;v^a_k, \;u^{\pm a}\}$:
\bea
\{x,\; v_k^a, \;u^{\pm a}\} & \Rightarrow & \{x,\; v_k^{\pm}, \;u^{\pm a}\}
\Rightarrow \{x,\;z^{++},z^{--},z^0,\;w^{+a},w^{-a}\} \label{5.9.1}
\eea
\bea
z^{++} &=& {v^{++}\over v^{-+}}\;,\;z^{--} = v^{-+}v^{--}\;, \;
z^0 = v^{-+}\;, \; v^{\pm\pm} = u^{\pm}_k v^{\pm k} \nn \\
v^{-+} &=& u^-_k v^{+k}\;, \;v^{+-} = u^+_kv^{-k}\;, \;
v^{++}v^{--}-v^{-+}v^{+-} = 1 \label{5.9.2} \\
w^+_a &=& u^+_a-z^{++}u^-_a\;,\; w^-_a = u^-_a\;, \; v^+_k = -z^0w^+_k\;, \;
v^-_k =  -{1\over z^0}(w^-_k + z^{--}w^+_k)\;. \label{5.9.3}
\eea
It is instructive to see how the original local $Sp(1)$ group is
realised in this basis
\bea
\delta z^{++} &=& \tau^{ij}(x)w^+_iw^+_j \;\equiv\; {\hat \tau}^{++}(x,w)\;,
\;\delta z^{--} = {\hat \tau}^{--}+2\;{\hat \tau}^{+-}z^{--}\;,
\;\delta z^0 = {\hat \tau}^{+-}z^0 \nn \\
\delta w^+_a &=& -{\hat \tau}^{++} w^-_a, \;\;\; \delta w^-_a\;=\;0\;.
\label{5.9.4}
\eea
Comparing with the results of ref. \cite{{a10},{a11}},
one observes that the newly introduced harmonics $w^{\pm i}$ are
composed of $u^{\pm a}$ and behave under the local $Sp(1)$ group
precisely in the same manner as the harmonics $w^{\pm a}$ entering the general
$N=2$ matter
Lagrangian in a $N=2$ supergravity background are composed of the $N=2$
harmonic superspace $u$'s and behave under the transformations called in
\cite{a11} quaternionic (analogs of the K\"ahler \cite{a8} and hyper-K\"ahler
\cite{{a11},{a2}} transformations). The object $z^{++}$ stands for the
supergravity compensator $N^{++}$. As we will see later on, this analogy is
by no means accidental (see Sect. 10).

{}From \p{5.9.4} one concludes that $z^{++},\;z^{--},\;z^0$ transform
independently of each other, i.e. each transforms through itself, $x$
and  $w^{\pm a}$. So, one can extract from \p{5.9.1} invariant
subspaces containing smaller number of $z$'s and overlapping over the
harmonic part $\{w^{\pm a}\}$ which in its
own right constitutes an invariant subspace (together with $x$).
In particular,
there exist three invariant subspaces each including only one of the $z$
coordinates
\bea
(x, z^{++},z^{--},z^0, w^{\pm a}) \Rightarrow
\{ (x, z^{++}, w^{\pm a}), \; (x,z^{--}, w^{\pm a}), \;
(x,z^0, w^{\pm a})\} \Rightarrow \{(x, w^{\pm a})\} \label{5.9.5}
\eea

This structure of the bi-harmonic space manifests itself in the form
of the covariant derivatives in the analytic basis
\bea
 Z^{\;++} &=& (z^0)^2{\partial\over \partial z^{--}} \nn \\
Z^{\;--} &=& (z^0)^{-2} \{ z^{--} ( z^0{\partial\over \partial z^0}+
z^{--}{\partial\over \partial z^{--}}) -
{\partial\over \partial z^{++}}+\partial^{--}_w \}
\nn \\
Z^{\;0} &=& z^0{\partial\over \partial z^0} \label{5.10.1} \\
{\cal D}^{\;++} &=& \partial^{++}_w -z^{++}[{\cal D}^{\;0} -
 z^0{\partial\over \partial z^0}-
z^{++}{\partial\over \partial z^{++}}] +
[(z^0)^2-1]{\partial\over \partial z^{--}} \nn \\
{\cal D}^{\;--} &=& Z^{\;--} + {\partial\over \partial z^{++}} \nn \\
{\cal D}^{\;0} &=& \partial^{\;0}_{w} + 2\;
(z^{++} {\partial\over \partial z^{++}}
- z^{--}{\partial\over \partial z^{--}})\;.
\label{5.10.2}
\eea

Considering a general field on the bi-harmonic space, $\phi(x,z,w)$
(we discard possible external indices), we may impose on it
the following $Sp(1)$ covariant analyticity constraints
\bea
 Z^{\;++}\phi=0 & \Rightarrow & {\partial\over \partial z^{--}}\phi=0\;, \nn
\\
{\rm or}\;\;\; Z^{\;0}\phi=0 & \Rightarrow & {\partial\over
\partial z^{0}}\phi=0 \;,\nn \\
{\rm or}\;\;\; (Z^{\;--}-{\cal D}^{\;--})\;\phi=0
& \Rightarrow & {\partial\over \partial z^{++}}\phi=0\;, \label{5.10.3}
\eea
or any combination of them simultaneously. Indeed, any pair of the
corresponding differential operators is closed under commutation; this
follows also  from \p{5.8.3}, \p{5.8.a}.
The covariance of eqs. \p{5.10.3} with
respect to $Sp(1)$ transformations follows from the transformation
properties of the derivatives involved
\bea
\delta {\partial\over \partial z^{0}}=0\;,\;\;
\delta {\partial\over \partial z^{++}}=0\;, \;\;
\delta {\partial\over \partial z^{--}}=
-2 {\hat \tau}^{+-} {\partial\over \partial z^{--}}\;.\label{5.10.4}
\eea
The solutions of these constraints are the fields living on the aforementioned
invariant subspaces of the bi-harmonic space. In particular, three subspaces
indicated in eq. (3.15) correspond to imposing first and second, second and
third, first and third of the constraints (3.18), respectively. The
fields on the subspace $(x,w^{\pm a})$ are singled out by employing the
whole set of  these constraints.

Now we will show that an arbitrary field $f_{ijk\dots}(x)$
(having the standard
transformation properties \p{5.2.1} under local $Sp(1)$) can be rewritten as a
constrained
analytic field on the above bi-harmonic space. Correspondingly, the
group $Sp(1)$ is realised on such fields by the differential operators
$Z^{(ij)}$.

Converting indices $i,j,k,\ldots$ of the field $f_{ijk\ldots}(x)$ with
the harmonics $v_i^a$ we form a field which is inert under the local
$Sp(1)$ and transforms only under the rigid $SU(2)$ acting on the
indices $a,b,c,\ldots$
\bea
f_{ab\ldots c}\;(x,v)=v_a^i v_b^j\ldots v_c^k \;f_{ij\ldots k}\;(x)\;.
\label{5.11.1}
\eea
Now the $Sp(1)$ transformation of $f_{ij\ldots k}(x)$ is induced by
the transformation of the variables $v_a^i$ and so it is represented by
the differential operator $\Gamma^{(kl)}$ \p{5.7.2}
\bea
\delta^* f(x,v)\equiv f'(x,v)-f(x,v) =\tau_k^l(x)\Gamma^{(k}{}_{l)}f(x,v)
\;. \nn
\eea
Further, projecting  $f_{ab\ldots c}\;(x,v)$ onto the monomials
$( u^{+a}\ldots u^{+c})$ one defines
\bea
f^{+n}(x,w,z^0)&=&(-1)^n v^{+i_1}\ldots v^{+i_n}f_{i_1\ldots
i_n}(x)\nn \\
&=&(z^0)^n\; w^{+i_1}\ldots w^{+i_n}f_{i_1\ldots i_n}(x)\nn \\
&\equiv& (z^0)^n \;{\tilde f}^{+n}(x,w)\;. \label{5.11.2}
\eea
These functions satisfy the relations
\bea
{\cal D}^{\;++}f^{+n} &=& \partial^{\;++}_w\;f^{+n}\;=\;0 \nn \\
{\cal D}^{\;0}f^{+n} &=& nf^{+n}\nn \\
Z^{\;0} f^{+n} &=& nf^{+n}\nn \\
Z^{\;++}f^{+n} &=& 0 \nn \\
(Z^{--}-{\cal D}^{--})f^{+n} &=& {\partial\over \partial z^{++}}f^{+n}\;=\;0
 \label{5.12.1}
\eea
and transform in the following way under the local $Sp(1)$
\bea
\delta f^{+n}(x,w,z^0) &=& f^{+n'}(x',w',z^{0'})-f^{+n}(x,w,z^0)\;=\;0 \nn \\
\delta  {\tilde f}^{+n}(x,w) &=& {\tilde f}^{+n'}(x',w')- {\tilde f}^{+n}(x,w)
\;=\;-n{\hat \tau}^{+-} {\tilde f}^{+n}(x,w)\;.\nn
\eea

To identify the place which $f^{+n}$ occupy among general functions on
the bi-harmonic space, we give here a general characterisation of such
functions.

By analogy with the harmonic functions on the ordinary sphere $S^2$ we
restrict ourselves  to the following functions $f^{(p,q)}(z,w)$
(for brevity, we omit the $x$ dependence)
\bea
{\cal D}^{\;0} f^{(p,q)}(z,w) &=& q\;f^{(p,q)}(z,w) \label{5.12.2a} \\
Z^{\;0} f^{(p,q)}(z,w) &=& p\;f^{(p,q)}(z,w) \;,\label{5.12.2b}
\eea
which corresponds to considering invariant functions on  $S^2\times S^2$
$\equiv Sp(1)/U(1)\times SU(2)/U(1)$. In the basis $(z,w)$ the condition
\p{5.12.2b} is easy to solve: it simply states that $f^{(p,q)}(z,w)$
is a homogeneous function of degree $p$ with respect to $z^0$
\bea
f^{(p,q)}(z,w) = (z^0)^p\;{\tilde f}^{(p,q)}(z^{++},z^{--},w)\;, \;\;\;
\delta {\tilde f}^{(p,q)} = -p\;{\hat \tau}^{+-}{\tilde f}^{(p,q)}\;.
\label{5.12.3}
\eea

Using the analyticity conditions \p{5.10.3}, we may now covariantly
constrain, in one way or another, the dependence of $f^{(p,q)}(z,w)$ on
$z^{++},z^{--}$. The simplest possibility is to completely eliminate
this dependence
\bea
Z^{\;++}f^{(p,q)}_{(1)} = (Z^{\;--}-{\cal D}^{\;--})\;f^{(p,q)}_{(1)}=0
\;\Rightarrow \; {\tilde f}^{(p,q)}_{(1)}=
{\tilde f}^{(p,q)}_{(1)}(w)\;. \label{5.13.1}
\eea
However, as it follows from \p{5.8.3}, ordinary harmonic derivatives
${\cal D}^{\;++},\;{\cal D}^{\;--}$ commute not with all operators
defining the above analyticity. So, the result of action of these
derivatives on $f^{(p,q)}_{(1)}$ is not in general again the field of
the type \p{5.13.1}, e.g.
\bea
{\cal D}^{\;++}f^{(p,q)}_{(1)} &=&
[\;\partial^{\;++}_w + z^{++}(Z^0-{\cal D}^{\;0})\;]f^{(p,q)}_{(1)}(w,z) \nn\\
{\cal D}^{\;--}f^{(p,q)}_{(1)} &=&
(z^0)^{-2}[\;\partial^{\;--}_w + z^{--}Z^{\;0}\;]f^{(p,q)}_{(1)}\;. \nn
\eea
These fields are of the type \p{5.13.1} only provided $q=p=0$. One may
still achieve this property for ${\cal D}^{\;++}f^{(p,q)}_{(1)}$ under a
weaker condition
\bea
(Z^{\;0}-{\cal D}^{\;0})f^{(p,q)}_{(1)}=0\; \Rightarrow \; q=p\;.
\label{5.13.2}
\eea
In this case the field
\bea
{\cal D}^{\;++}f^{(q)}_{(1)}(w,z^{0})=\partial^{\;++}_w f^{(q)}_{(1)}(w,z^0)
\label{5.13.3}
\eea
is covariant under $Sp(1)$ transfromations, which can be checked
explicitly, using
\bea
\delta \partial^{\;++}_w=-{\hat \tau}^{++}(Z^{\;0}-{\cal D}^{\;0})-
2{\hat \tau}^{+-}{\partial\over\partial z^{--}}\;. \label{5.13.4}
\eea
We call the fields satisfying \p{5.13.1}, \p{5.13.2} the fields of
the first kind.
The functions $f^{+n}(w,z^0)$ considered before belong just to this class,
they are singled out by the additional covariant constraint
\bea
{\cal D}^{\;++}f^{(q)}_{(1)}(w,z^0) = \partial^{\;++}_w
f^{(q)}_{(1)}(w,z^0) = 0 \;\Rightarrow \;
f^{(q)}_{(1)}(w,z^0) = (z^0)^q\; w^{+i_1}\ldots
w^{+i_q}f_{i_1\ldots i_q}\;. \nn
\eea

Another interesting class of functions having a restricted
$z^{++},z^{--}$ dependence is defined by the constraints
\bea
(Z^{\;0}-{\cal D}^{\;0})\;f^{(p,q)}_{(2)} &=& 0 \;\Rightarrow \;
 f^{(p,q)}_{(2)}\equiv f^{(q)}_{(2)}\label{5.14.1a}\\
({\cal D}^{\;--} - Z^{\;--})\;f^{(p,q)}_{(2)} =
{\partial\over\partial z^{++}}f^{(q)}_{(2)} &=& 0 \nn \\
(Z^{\;++}-{\cal D}^{\;++})f^{(q)}_{(2)}\;=\;({\partial\over\partial z^{--}}
- \partial^{\;++}_{w})\;f^{(q)}_{(2)} &=& 0\;. \label{5.14.1b}
\eea
These fields do not depend on $z^{++}$, while their $z^{--}$ dependence
is restricted by
\bea
{\partial\over\partial z^{--}}\;f^{(q)}_{(2)}=\partial^{++}_wf^{(q)}_{(2)}\;,
\label{5.14.2}
\eea
so any such field is actually a collection of all the degrees of the
harmonic derivative $\partial^{++}_w$ of the lowest component in its
$z^{--}$ expansion. So, we may limit our
consideration to the component ${\tilde f}^{(q)}_{(2)}\Vert$ where the
vertical line means the value at $z^{--}=0$. It transform as follows
\bea
\delta {\tilde f}^{(q)}_{(2)}\Vert =
 -q\;{\hat \tau}^{+-}\; {\tilde f}^{(q)}_{(2)}\Vert
 -{\hat \tau}^{--}\;\partial^{\;++}_w {\tilde f}^{(q)}_{(2)}\Vert\;.\nn
\eea

The important difference of the second kind fields from the first
kind ones consists in that $\partial^{\;++}_w {\tilde f}^{(q)}_{(2)}$ is
again the field of the second kind, its $Sp(1)$ weight and $U(1)$
charge being equal to $q+2$,
\bea
\delta (\partial^{\;++}_w {\tilde f}^{(q)}_{(2)})\Vert =
 -(q+2)\;{\hat \tau}^{+-}(\partial^{\;++}_w {\tilde f}^{(q)}_{(2)})\Vert
-{\hat \tau}^{--}\;\partial^{\;++}_w \;
(\partial^{\;++}_w {\tilde f}^{(q)}_{(2)})\Vert\;, \nn
\eea
while $\partial^{++}_w {\tilde f}^{(q)}_{(1)}$ is a field of a new
type with the nonaltered  $Sp(1)$ weight $q$ and the $U(1)$
charge shifted by $2$ :
\bea
\delta (\partial^{++}_w {\tilde f}^{(q)}_{(1)})=
-q\;{\hat \tau}^{+-}\;(\partial^{++}_w {\tilde f}^{(q)}_{(1)})\;.\nn
\eea
(it belongs to the type (3.26)).

Concluding this Section, we comment on several points which are
important for understanding what we will do in the sequel.

As a first remark we point out that the bi-harmonic space $\{x,v,u\}$
contains evident subspaces $\{x,v^+,v^-\}$ and $\{x,u^+,u^-\}$ which
are trivially closed both under the local $Sp(1)$ and rigid $SU(2)$
transformations. As we will see later, these subspaces bear no
direct relevance to the quaternionic geometry constraints, in
contradistinction to the subspace $\{x,w^+,w^-\}$. Just the latter
is preserved by these
constraints, thus expressing most clearly the concept of harmonic
analyticity in application to quaternionic geometry.

We also note that the functions given on the bi-harmonic space are in
general represented by series which are not a naive double
harmonic expansion with respect to the sets of harmonics $v^{\pm i}$
and $u^{\pm a}$. E.g., let us consider a function
$f^{(q)}(x,w^+,w^-)$. It has a standard $w^{\pm i}$-expansion
\bea
f^{(q)}(x,w^+,w^-)=\sum f^{(i_1\ldots i_nj_{n+1}\ldots j_{n-q)}}(x)\;
w^+_{i_1}\ldots w^+_{i_n}w^-_{j_{n+1}}\ldots w^-_{j_{n-q}}\;.\nn
\eea
However, in the basis $(v^{\pm i},u^{\pm a})$ it is
\bea
f^{(q)}(x,w) = {\tilde f}^{(q)}(x,v,u) =
\sum f^{(i_1\ldots i_nj_{n+1}\ldots j_{n-q)}}(x)\;
v^+_{i_1}\ldots v^+_{i_n}u^-_{j_{n+1}}\ldots u^-_{j_{n-q}}(-v^{-+})^{-n}\;.
\nn
\eea
The presence of unusual factors $(v^{-+})^{-n}$ is necessary for ensuring
the property
\bea
Z^{\;0} f^{(q)}=(v^+{\partial\over\partial v^+} -
 v^-{\partial\over\partial v^-})f^{(q)}=0
\label{5.15.1}
\eea
in the basis $(v,u)$ (in the basis $(z,w)$ this property is evident).
Clearly, \p{5.15.1} can never be achieved within a naive double
$(v,u)$ harmonic expansion.

Having  at our disposal the object $v^{-+}$
with the
vanishing $U(1)$ charge but with the unit $Sp(1)$ weight, we are to
take caution while transferring into the bi-harmonic space some
statements valid in the ordinary harmonic space. In particular, the
constraints
\bea
Z^{\;++}f^{(p,q)} &=& v^+{\partial\over\partial v^-}\;f^{(p,q)}\;=\;0 \nn \\
Z^{\;0}f^{(p,q)} &=& (v^+{\partial\over\partial v^+} -
v^-{\partial\over\partial v^-}) f^{(p,q)}\;=\;p\;f^{(p,q)}\nn
\eea
do not imply that the  $v^{-+}$ expansion of $f^{(p,q)}$ terminates.
In parallel with the standard term
\bea
v^{+i_1}\ldots v^{+i_p}f^{(q-p)}(x,u) \nn
\eea
one is led to include into the $v$-decomposition of $f^{(p,q)}$  an
infinite collection of the terms of the form
\bea
v^{+i_1}\ldots v^{+i_p}v^{+i_{p+1}}\ldots v^{+i_{p+n}}f^{(q-p-n)}(x,u)
\;(v^{-+})^{-n} \nn
\eea
with $n\geq 1$. Each term of that sort is annihilated by $Z^{++}$ and
carries the $Sp(1)$ weight $p$.

 In what follows, to avoid subtleties of the bi-harmopnic
decompositions, we will basically use the parametrization of the
harmonic part of the bi-harmonic space by the coordinates $w^{\pm i},
z^{++}, z^{--}, z^0$. We will assume standard harmonic expansions with
respect to $w^{\pm i}$, not specifying the dependence on $z^{++},
z^{--}$, except for a few special cases, including the important case
when there is no such a dependence at all (as was already mentioned,
the standard fields \p{5.2.1} belong just to this latter class).
Recall that the $z^0$ dependence of the bi-harmonic fields is fixed as
in eq. \p{5.11.2}, as soon as we limit our consideration to the fields
with a definite  $Sp(1)$ weight.

\setcounter{equation}{0}
\section{Quaternionic geometry constraints as the integrability
conditions for bi-harmonic analyticity}

Here we show that the defining constraint \p{5.5.2}, being rewritten in
the bi-harmonic extension of the original $4n$ dimensional manifold $M$,
can be given a natural interpretation of the integrability conditions
for the existence of special analytic functions on this extended
manifold.

Thus we deal with the  bi-harmonic space
\bea
\{x^{\mu i}, \; v_a{}^l, \; u^{\pm}_a \} \label{5.17.1}
\eea
and treat ordinary fields \p{5.2.1} defined originally as
unconstrained functions on $\{ x^{\mu i} \}$ as fields \p{5.11.2} on the
enlarged space \p{5.17.1}. They may carry any number of external $Sp(1)$
indices and an arbitrary $Sp(1)$ weight (equal to their $U(1)$
charge). Their content is the same as that of \p{5.2.1} provided they
satisfy the constraints \p{5.12.1}. Of course, one is at liberty to consider
more general functions on \p{5.17.1}, in accordance with the analysis in the
previous Section.

The main advantage of such an approach consists in the possibility to realize
the generators of the tangent space local $Sp(1)$ by differential operators.
An equivalent representation of the covariant derivative \p{5.3.1} in the
space \p{5.17.1} is given by
\bea
{\cal D}_{\alpha a}=v_a{}^ie^{\mu k}_{\alpha i}\partial_{\mu k}-
 v_a{}^i \omega_{\alpha i(\sigma\rho)}\Gamma^{(\sigma\rho)}
 - {1\over 2} v_a{}^i\omega_{\alpha i(lk)}
(v_b{}^l\epsilon^{kt}{\partial\over\partial v_b{}^t} +
v_b{}^k\epsilon^{lt}{\partial\over\partial v_b{}^t})\;. \label{5.18.1}
\eea
One easily checks that the action of \p{5.18.1} on the field of the type
\p{5.11.1} yields the correct covariant derivative of the coefficients
$f_{ij\ldots k}(x)$
\bea
{\cal D}_{\alpha a}(v_{a_1}{}^{i_1}v_{a_2}{}^{i_2}\ldots v_{a_n}{}^{i_n}
f_{i_1i_2\ldots i_n}(x)) =
v_{a}{}^{i}v_{a_1}{}^{i_1}v_{a_2}{}^{i_2}\ldots v_{a_n}{}^{i_n}
{\cal D}_{\alpha i}f_{i_1i_2\ldots i_n}(x)\;. \label{5.18.2}
\eea
It is easy to compute the commutator of two derivatives \p{5.18.1}
\bea
[{\cal D}_{\alpha a},{\cal D}_{\beta b}]_\rho{}^{\rho'}=
2\;\delta_\rho{}^{\rho'}
\Omega_{\alpha\beta}RZ_{(ab)} +
\epsilon_{ab}\{{\phantom q}\}_{(\alpha\beta)\rho}{}^{\rho'}\;,
\label{5.18.3}
\eea
where the braces stand for the $Sp(n)$ curvature terms and $Z_{(ab)}$ is the
covariant derivative with respect to the coordinates $v_{a}{}^{i}$ ( it is
defined in eqs. \p{5.7.3}). Now, the basic constraint \p{5.5.2} is written as
\bea
[{\cal D}_{\alpha (a},{\cal D}_{\beta b)}]_\rho{}^{\rho'}=
2\;\delta_\rho{}^{\rho'}
\Omega_{\alpha\beta}RZ_{(ab)}\;.
\label{5.19.1}
\eea

{\it Digression}. This is a good place to comment on the twistor space.
Harmonics $v^{\ l}_a$ generate the sphere $S^3$ at each point of the
quaternionic manifold $M$. Factorizing by $U(1)$ group acting on harmonics
$v^l_a$ as $v_1^{\ l}\mapsto \exp (i\phi )v_1^{\ l}$,
$v_2^{\ l}\mapsto \exp (-i\phi )v_2^{\ l}$ we obtain the space $S^2$
parametrized by complex
structures at the point of $M$: the complex structure corresponding to
$v_a^{\ l}$ is
\be I_{\alpha i}^{\ \ \ \beta j}=(v_{1i}v_2^{\ j}
+v_{2i}v_1^{\ j})\delta^\beta_\alpha\ .\ee
The space of all these complex structures attached at each
point of $M$ is called the twistor space ${\bf Z}$ in \cite{a12}.
We conclude that the space
$\{ x^{\mu i},v_a^{\ l}\}$ with the above equivalence relation is
the twistor space of $M$.

The antiholomorphic direction with respect to the natural complex
structure on $S^2$ is spanned by $Z_{11}$. We define now vector fields
\be X_\alpha =v_1^{\ i} e^{\mu k}_{\alpha i}\partial_{\mu k}-\frac{1}{2}
v_1^{\ i} \omega_{\alpha i(lk)}(v_b^{\ l}\epsilon^{kt}
\frac{\partial}{\partial v_b^{\ t}}+v_b^{\ k}\epsilon^{lt}
\frac{\partial}{\partial v_b^{\ t}})\ .\ee
Using \p{5.19.1} one gets
\be [X_\alpha ,X_\beta ]=2\Omega_{\alpha\beta}RZ_{11}+v_1^{\ i}
(\omega_{\beta i\alpha}^{\ \ \ \ \gamma}-\omega_{\alpha i\beta}^{\ \ \
\ \gamma})X^\gamma\ .\label{ha}\ee
We have also $[Z_{11},X_\alpha ]=0$. Define now a tensor of an almost
complex structure on the twistor space to have the space spanned by
$\{ X_\alpha ,Z_{11}\}$ as its eigenspace with the eigenvalue $i$. The
above arguments imply its integrability. This is the integrable
complex structure discussed in \cite{a12}.

In \cite{a12} for $R\neq 0$ a complex contact structure on ${\bf
Z}$ was introduced. It can be also easily read off the equation
(\ref{ha}). Let $\sigma$ be a local one-form satisfying $(\sigma ,
Z_{11})=1$ and $(\sigma ,X_\alpha)=0$ for all $\alpha$, where $(\sigma, X)$ is
the value of the one-form $\sigma$ on a vector field $X$. The form $\sigma$
defines a complex contact structure on $Z$. To show this, one can write the
explicit expression for $\sigma$ in the local coordinates. However, it is
enough to use the identity
$(d\sigma ,X_\alpha\wedge X_\beta )=-(\sigma ,[X_\alpha ,X_\beta ])$.
We find using (\ref{ha}) that
\be (d\sigma ,X_\alpha \wedge X_\beta )=-2\Omega_{\alpha\beta}R \ .
\label{haha}\ee
The kernel of
$\sigma$ is spanned by $X_\alpha$. For $R\neq 0$ equation
(\ref{haha}) shows that $d\sigma$ is nondegenerate on the kernel of
$\sigma$. Therefore, $\sigma$ is a well-defined complex contact one-form.

Now we return to the main line of our exposition.
What we have gained at this step is that the local $Sp(1)$ indices
$i,j,\ldots$ have been replaced by the rigid $SU(2)$ ones $a,b,\ldots $ and
the quaternionic geometry constraint has been reformulated entirely in terms
of appropriate covariant derivatives. Now we may proceed in the way well
known from the previous
examples of utilyzing the harmonic analyticity \cite{{a1},{a2},{a3},{a4}}:
multiply \p{5.19.1} by the extra harmonics $u^{+a},u^{+b}$ and recast
this relation into the form of the integrability condition
\bea
[{\cal D}_\alpha^{\;+} ,{\cal D}_\beta^{\;+}]= - 2\;\delta_\rho{}^{\rho'}
\Omega_{\alpha\beta}RZ^{\;++},\;\;\;{\cal D}_\alpha^{\;+}=
u^{+a}{\cal D}_{\alpha a},\;\;\;
Z^{\;++}=-u^{+a}u^{+b}Z_{(ab)}\;.
\label{5.19.2}
\eea
The full set of commutators in this new representation can be restored after
introducing
\bea
{\cal D}_\alpha^{\;-}=
u^{-a}{\cal D}_{\alpha a}=[\partial^{\;--},{\cal D}_\alpha^{\;+}]\;.
\label{5.19.3}
\eea
Then the remaining commutators are
\bea
[{\cal D}_{\alpha}^{\;+},{\cal D}_{\beta}^{\;-}]_\rho{}^{\rho'} =
\;\delta_\rho{}^{\rho'}
\Omega_{\alpha\beta}RZ^{\;0} + \{{\phantom q}
\}_{(\alpha\beta)\rho}{}^{\rho'},\;\;\;
[{\cal D}_{\alpha}^{\;-},{\cal D}_{\beta}^{\;-}]_\rho{}^{\rho'} =
2\;\delta_\rho{}^{\rho'}
\Omega_{\alpha\beta}RZ^{\;--}\;.
\label{5.20.1}
\eea
One should also add the commutators between the harmonic derivatives \p{5.8.3}
and the commutators of the latter with ${\cal D}_{\alpha}^{\;+}, \;
{\cal D}_{\beta}^{\;-}$:
they immediately follow from the definition of ${\cal D}_{\alpha}^{\;\pm}$.

The most important step is to single out the
minimal closed set of operators which forms a Cauchi-Riemann (CR)-structure
and is equivalent to the quaternionic geometry constraints \p{5.5.2}.
In the present case
this set includes the following operators
\bea
\{ {\cal D}_{\alpha}^{\;+}, Z^{\;++}, Z^{\;0}, {\cal D}^{\;++}, {\cal D}^{\;0},
{\cal D}^{\;--}-Z^{\;--} \}\;.
\label{5.20.2}
\eea
Their algebra is given by
\bea
[{\cal D}_\alpha^{\;+}, {\cal D}_\beta^{\;+}] = - 2\;\delta_\rho{}^{\rho'}
\Omega_{\alpha\beta}RZ^{\;++} \label{5.20.3a}
\eea
\bea
[{\cal D}^{\;++},{\cal D}_\alpha^{\;+}] = 0 \label{5.20.3b}
\eea
\bea
[Z^{\;++},{\cal D}_\alpha^{\;+}] = 0\;, \;\;
[{\cal D}^{\;--}-Z^{\;--},{\cal D}_\alpha^{\;+}]=0
\label{5.20.3c}
\eea
\bea
[{\cal D}^{\;0},{\cal D}_\alpha^{\;+}] = [Z^{\;0},{\cal
D}_\alpha^{\;+}]={\cal D}_\alpha^{\;+} \label{5.20.3d}
\eea
and also
\bea
[{\cal D}^{\;++},Z^{\;++}] = 0\;,\;\;
[{\cal D}^{\;++},{\cal D}^{\;0}] = -2\;{\cal D}^{\;++} \nn
\eea
\bea
[{\cal D}^{\;0},Z^{\;++}] = 2\;Z^{\;++}\;, \;\;
[Z^{\;0},{\cal D}^{\;++}] = 2\;Z^{\;++}\;,\;\;
[Z^{\;0},Z^{\;++}] = 2\;Z^{\;++} \nn
\eea
\bea
[{\cal D}^{\;--}-Z^{\;--},{\cal D}^{\;0}] =
-2\;({\cal D}^{\;--}-Z^{\;--})\;, \;\;
[{\cal D}^{\;--}-Z^{\;--},Z^{\;0}] = 0 \nn
\eea
\bea
[{\cal D}^{\;--}-Z^{\;--},{\cal D}^{\;++}]=
Z^{\;0}-{\cal D}^{\;0}\;,\;\; [{\cal D}^{\;--}-Z^{\;--},Z^{\;++}]=0\;.
\label{5.21.1}
\eea
The relations \p{5.21.1} are written down for
completeness, these are satisfied
in the central basis by the definition of harmonic derivatives. The basic
relations which actually underly the quaternionic geometry are those given
in eqs. \p{5.20.3a} - \p{5.20.3d}. To be convinced of that
these relations indeed
amount to the original constraint \p{5.5.2} one proceeds as follows.

First, eqs. \p{5.20.3d} imply that ${\cal D}_\alpha^{\;+}$
has coinciding $U(1)$ charge
(the eigenvalue of ${\cal D}^{\;0}$) and $Sp(1)$ weight
(the eigenvalue of $Z^{\;0}$).
So, its most general form consistent with \p{5.20.3d} and the choice of
$Sp(n)$ as the tangent space group
is given by (in the basis $\{ z,w \}$)
\bea
{\cal D}^+_\alpha= E^{+\mu k}_\alpha \partial_{\mu k}
+\omega^+_{\alpha(\rho\sigma)}\Gamma^{(\rho\sigma)}
+\omega ^-_\alpha Z^{++} +\omega^+_\alpha Z^0- \omega ^{+3}_\alpha Z^{--}\;.
\label{5.21.2}
\eea
Here
\bea
Z^{\;0} E^{\;+\mu k}_\alpha = E^{\;+\mu k}_\alpha  &\Rightarrow &
 E^{\;+\mu k}_\alpha =z^0\;{\tilde E}^{\;+\mu k}_\alpha \nn \\
Z^{\;0}\omega^+_{\alpha(\rho\sigma)} = \omega^+_{\alpha(\rho\sigma)}
&\Rightarrow &
\omega^+_{\alpha(\rho\sigma)}=z^0\;{\tilde
\omega}^+_{\alpha(\rho\sigma)} \nn \\
Z^{\;0} \omega^-_{\alpha} = -\omega^-_{\alpha} &\Rightarrow &
\omega^-_{\alpha}=(z^0)^{-1}\;{\tilde \omega}^-_{\alpha} \nn \\
Z^{\;0}\omega^+_{\alpha} = \omega^+_{\alpha} &\Rightarrow &
\omega^+_{\alpha}=z^0\;{\tilde \omega}^+_{\alpha} \nn \\
Z^{\;0}\omega^{+3}_{\alpha} = 3\;\omega^{+3}_{\alpha} &\Rightarrow &
\omega^{+3}_{\alpha}=z^0\;{\tilde \omega}^{+3}_{\alpha} \label{5.21.3}
\eea
and the objects with tilde display no $z^0$ dependence. Then,
exploiting \p{5.20.3c} and the commutation relations between
$Z$-operators, one completely fixes the $z$ -dependence of these objects
\bea
({\cal D}^{\;--}-Z^{\;--})\;{\tilde E}^{+\mu k}_\alpha =
Z^{\;++}\;{\tilde E}^{+\mu k}_\alpha =0 &\Rightarrow &
{\tilde E}^{+\mu k}_\alpha = {\tilde E}^{+\mu k}_\alpha (x,w) \nn \\
({\cal D}^{\;--}-Z^{\;--})\;{\tilde \omega}^+_{\alpha(\rho\sigma)} =
Z^{\;++}\; {\tilde\omega}^+_{\alpha(\rho\sigma)} = 0 &\Rightarrow &
{\tilde\omega}^+_{\alpha(\rho\sigma)}=
{\tilde\omega}^+_{\alpha(\rho\sigma)}(x,w) \nn \\
({\cal D}^{\;--}-Z^{\;--})\;{\tilde \omega}^{+3}_\alpha =
Z^{\;++}\;{\tilde \omega}^{+3}_\alpha = 0 &\Rightarrow &
{\tilde \omega}^{+3}_\alpha={\tilde
\omega}^{+3}_\alpha(x,w) \nn \\
({\cal D}^{\;--}-Z^{\;--})\;{\tilde \omega}^{+}_\alpha = 0\;,\;\;
{\partial\over\partial z^{--}}\;{\tilde \omega}^{+}_\alpha=
{\tilde \omega}^{+3}_\alpha
&\Rightarrow &  \nn \\
{\tilde \omega}^{+}_\alpha={\tilde
\omega}^{+}_\alpha(x,w) &+& z^{--}\;{\tilde\omega}^{+3}_\alpha(x,w) \nn \\
({\cal D}^{\;--}-Z^{\;--})\;{\tilde \omega}^{-}_\alpha = 0\;,\;\;
{\partial\over\partial z^{--}}\;{\tilde \omega}^{-}_\alpha=
2\;{\tilde \omega}^{+}_\alpha &\Rightarrow & \nn \\
{\tilde \omega}^{-}_\alpha = {\tilde
\omega}^{-}_\alpha(x,w) + 2\;z^{--}\;{\tilde\omega}^{+}_\alpha(x,w)
&+&(z^{--})^2\;{\tilde\omega}^{+3}_\alpha(x,w)\;.\label{5.22.1}
\eea
Finally, by using \p{5.20.3b}, we fix the $w$-dependence of
${\tilde E}^{+\mu k}_\alpha,\; {\tilde\omega}^+_{\alpha(\rho\sigma)}, \;
{\tilde \omega}^{-}_\alpha,\;{\tilde \omega}^{+}_\alpha$ and
${\tilde \omega}^{+3}_\alpha$ as
\bea
{\tilde E}^{+\mu k}_\alpha &=&  w^{+i} e^{\mu k}_{\alpha i}(x)\;,\;\;
{\tilde\omega}^+_{\alpha(\rho\sigma)}=
 w^{+i}\omega_{\alpha i(\rho\sigma)}(x) \nn \\
{\tilde\omega}^{+3}_\alpha &=& - w^{+i}w^{+j}w^{+k}\omega_{\alpha(ijk)}(x)
\nn \\
{\tilde \omega}^{+}_\alpha &=& - w^{+i}\omega_{\alpha
i}(x)- w^{+i}w^{+j}w^{-k}\omega_{\alpha(ijk)}(x) \nn \\
{\tilde
\omega}^{-}_\alpha &=& -2\; w^{-i}\omega_{\alpha i}(x)-
w^{+i}w^{-j}w^{-k}\omega_{\alpha(ijk)}(x)\;, \label{5.22.2}
\eea
where the sign
``$ - $'' was chosen for further convenience. Combining  $\omega_{\alpha
i}(x)$ and $\omega_{\alpha(ijk)}(x)$ into the single object  \bea
\omega_{\alpha i(jk)}(x)\equiv \omega_{\alpha(ijk)}(x)+
\epsilon_{ij}\;\omega_{\alpha k}(x) +\epsilon_{ik}\;\omega_{\alpha j}(x)\;,
\nn  \eea
we
eventually represent ${\cal D}^{\;+}_\alpha$ \p{5.21.2} as \bea  {\cal
D}^{\;+}_\alpha &=& E^{+\mu k}_\alpha \partial_{\mu k}-
\omega^+_{\alpha(\rho\sigma)}\Gamma^{(\rho\sigma)}+
\omega^{+3}_\alpha\partial^{\;--}_w -\omega^{+-+}_\alpha Z^{\;0} \nn \\
&-& (\omega^{+--}_\alpha {\partial\over\partial z^{--}}
+2\;\omega^{+-+}_\alpha z^{--}{\partial\over\partial z^{--}}
+\omega^{+3}_\alpha {\partial\over\partial z^{++}})  \nn \\ &\equiv&
{\tilde{\cal D}}^{\;+}_\alpha  -z^0\;({\tilde\omega}^{+--}_\alpha
{\partial\over\partial z^{--}} +2\;{\tilde\omega}^{+-+}_\alpha
z^{--}{\partial\over\partial z^{--}} +{\tilde\omega}^{+3}_\alpha
{\partial\over\partial z^{++}}) \label{5.22.3} \\
{\tilde{\cal D}}^{\;+}_\alpha&=& z^0\;{\tilde{\nabla}}^{\;+}_\alpha -
z^0\;{\tilde\omega}^{+-+}_\alpha Z^{\;0} \nn \\
& \equiv& z^0\;(\tilde{\Delta}^{\;+}_\alpha -
{\tilde\omega}^{+}\Gamma +{\tilde\omega}^{+3}\partial^{\;--}_w) -
z^0\;{\tilde\omega}^{+-+}_\alpha Z^{\;0} \label{5.22.4}
\eea
$$[{\tilde{\cal D}}^{\;+}_\alpha, {\tilde{\cal D}}^{\;+}_\beta] = 0\;. $$
Here
\bea
\tilde{\Delta}^{\;+}_\alpha &=& {\tilde E}^{+\mu k}_\alpha \partial_{\mu k}=
w^{+i}e^{\mu k}_{\alpha i}(x) \partial_{\mu k} \nn \\
{\tilde\omega}^+_{\alpha(\rho\sigma)} &=&
w^{+i}\omega_{\alpha i(\rho\sigma)}(x) \nn \\
{\tilde\omega}^{+-+}_\alpha &=& w^{+i}w^{-j}w^{+k}\omega_{\alpha i(jk)}(x)
\nn\\
{\tilde\omega}^{+--}_\alpha &=& w^{+i}w^{-j}w^{-k}\omega_{\alpha i(jk)}(x)
\nn\\
{\tilde\omega}^{+3}_\alpha &=& w^{+i}w^{+j}w^{_k}\omega_{\alpha i(jk)}(x)\;.
\label{5.22.5}
\eea
Now, returning to the basis $\{v_a{}^i, u^{\pm a} \}$, it is easy to show
that
\be
{\cal D}^{\;+}_\alpha=u^{+a}{\cal D}_{\alpha a}\;, \label{5.22.6}
\ee
where ${\cal D}_{\alpha a} $is given by eq.\p{5.18.1}. Finally,
substituting \p{5.22.6} into eq.\p{5.20.3a} and keeping in mind that
$Z^{\;++}=-{1\over 2}u^{+a}u^{+b}Z_{ab}$, one arrives at the relation
\p{5.19.1}  which, as we have seen before, is the same as the original
constraint \p{5.5.2}.

Two comments are in order here.

Firstly, it is just the equality of the  $U(1)$
charge and the $Sp(1)$ weight of ${\cal D}^+_\alpha$ (eqs. \p{5.20.3d})
that guarantees the compatibility of the constraints
\p{5.20.3b} and the second of the constraints \p{5.20.3c}. Commuting
${\cal D}^{\;++}$ with the l.h.s. of the latter equation one gets
\be
[ {\cal D}^{\;0}-Z^{\;0}, {\cal D}^{\;+}_\alpha ]=0
\ee
that is fulfilled as a consequence of eqs.  \p{5.20.3d}.

Secondly, it should be stressed that the $SU(2)$ algebra for $Z_{ab}$
immediately
stems from the important relations contained in the set of the harmonic
constraints \p{5.21.1}
\bea
[{\cal D}^{\;++},Z^{\;++}]  = 0\;, \;\;[{\cal D}^{\;++},Z^{\;0}] = -
2\;Z^{\;++}\;,\;\;
[ Z^{\;0},Z^{\;++}] = 2\;Z^{\;++}\;.\label{5.23.0c} \eea
First of these equations implies $Z^{++}= -{1\over 2}u^{+a}u^{+b}Z_{ab}$,
then from the second one it follows  $Z^{\;0} =-u^{+a}u^{-b}Z_{ab}$ and,
finally,
substituting these expressions for $Z^{\;0}$ and $Z^{\;++}$ into the third
equation and using
$u^\pm_a$ algebra, one deduces the $Sp(1)$ commutation relations \p{5.7.4}.

To summarize, we have demonstrated that the defining constraints of the
quaternionic geometry  \p{5.5.2} in the space $\{x^{\mu k}\}$ amount
to the constraints \p{5.20.3a}-\p{5.20.3d} in the bi-harmonic space
$\{x^{\mu k}, \;v^{\pm i}, \;u^{\pm a} \}$=$\{x^{\mu k}, \;z,\;w \}$.
Like in all other cases to which the harmonic analyticity is relevant,
the property that the
operators $\{{\cal D}^{\;+}_\alpha,\; Z^{\;++}, {\cal D}^{\;0},\; Z^{\;0}, \;
{\cal D}^{\;--}-Z^{\;--}\}$
form a closed algebra amounts to the possibility to define covariantly
analytic
fields. This time, they are defined by the following analyticity conditions
\bea
{\cal D}^{\;+}_\alpha \Phi^{(p,q)}(x,v,u) &=& 0 \label{5.23.1a}\\
 Z^{\;++}\Phi^{(p,q)}(x,v,u) &=& 0 \label{5.23.1b}\\
({\cal D}^{\;--}-Z^{\;--})\Phi^{(p,q)}(x,v,u) &=& 0 \label{5.23.1c}\\
Z^{\;0}\Phi^{(p,q)}(x,v,u) &=& p\;\Phi^{(p,q)}(x,v,u) \label{5.23.1d}\\
{\cal D}^{\;0}\Phi^{(p,q)}(x,v,u) &=& q\;\Phi^{(p,q)}(x,v,u) \;.
\label{5.23.1e}
\eea
Eq.\p{5.23.1b} combined with the relations $[{\cal D}^{\;++},Z^{\;++}]=0\;, $
$ [{\cal D}^{\;++},Z^{\;0}]=-2\;Z^{++}$ imply that the harmonic derivative
${\cal D}^{\;++}$ preserves this analyticity provided $p=q$:
${\cal D}^{\;++}\Phi^{(q,q)}$
is again an analytic field with the $U(1)$ charge $q+2$ and the $Sp(1)$
weight $q$. Recall that the original tensor fields with the transformation law
\p{5.2.1} form a subclass in the variety of bi-harmonic fields $\Phi^{(q,q)}$.

Note that the conditions \p{5.23.1b} - \p{5.23.1e} mean that the
field $\Phi^{(p,q)}$ does not depend on the coordinates $z^{++}, z^{--}$ and
involves the factorized $z^{\;0}$ dependence.
So, one is eventually left with the
following harmonic analyticity condition (see definition \p{5.22.4})
\bea
(\tilde{\nabla}^{\;+}_\alpha - p\;\tilde{\omega}^{+-+}_{\alpha})\;
\tilde{\Phi}^{(p,q)}(x,w^\pm)=0\;,\;\;\;
\Phi^{(p,q)}(x,v,u) = (z^{0})^{p}\tilde{\Phi}^{(p,q)}(x,w)\;.
\label{5.24.1}
\eea
It implies, like in the HK case \cite{a2}, the existence of an invariant
analytic
subspace $\sim \{x^+_A, w^\pm_A\}$ in $\{x^{\mu i}, w^\pm \}$. The principal
difference from the HK case is that the harmonics $ w^\pm$ (and their
analytic
basis counterparts  $ w^\pm_A$) have non-trivial transformation properties
under the tangent space local $Sp(1)$ group (correspondingly,
$\tilde{\nabla}^{\;+}_{\alpha}$ in eq. \p{5.24.1} contains a differentiation
with respect to $w$). This peculiarity of the
quaternionic case will entail, in particular, the necessity of an additional
bridge while passing to the analytic world, namely the one relating the
$\tau$ - and $\lambda$ - basis   $w$ harmonics (see below).

For our further purposes it will be crucial that the analyticity
\p{5.23.1a} - \p{5.23.1e} implies, like in the HK and self-dual Yang-Mills
cases
\cite{{a1},{a2}}, the
existence of some basis and frame (the $\lambda$-world)
where this analyticity is
manifest. Just as in these cases, $\lambda$-world turns out to be most
appropriate for solving the integrability conditions
\p{5.20.3a} - \p{5.20.3d}, \p{5.21.1}. On all stages of computation we will
keep
the derivatives with respect to $z$-coordinates. This guarantees that we
will
not lose any information encoded in eqs. \p{5.20.3a} - \p{5.20.3d}, \p{5.21.1}.

\setcounter{equation}{0}
\section{Bridges to $\lambda$-world}

As was just mentioned, the existence of the CR-structure
\p{5.20.3a} - \p{5.20.3d}, \p{5.21.1} implies possiblity to pass to
the basis and frame called $\lambda$-world where the analyticity associated
with this structure becomes manifest. This means that
differential operators forming the CR-structure
are reduced in the $\lambda$-world to partial derivatives.

In principal, there are different choices of the change
of the coordinates $\{x^{\mu i},z^{(ab)},w^\pm\}$ leading to the analytic
basis. The minimal and most convenient option is as follows
\bea
z^{++}_A &=& z^{++}+v^{++}(x,w)\nn \\
z^{--}_A &=& t^2(x,w)\;[z^{--}+v^{--}(x,w)] \nn \\
z^0_A &=& t(x,w)\;z^0 \label{5.25.1}
\eea
\bea
w^{-i}_A &=& w^{-i} \nn \\
w^{+i}_A &=& w^{+i} - +v^{++}(x,w)w^{-i}\nn \\
x^{\pm\mu}_A &=& x^{\mu i}w^\pm_{i}+v^{\pm\mu}(x,w)\;. \label{5.25.2}
\eea
The bridges can be consistently chosen independent of the coordinates
$z$ in view of the bi-harmonic integrability conditions \p{5.21.1}. Then
the basic
integrability condition \p{5.20.3a} allows one to define the bridges by the
following constraints (analogous to those employed in the HK case \cite{a2})
\bea
{\cal D}^{\;+}_\alpha z^{++}_A=0 &\Rightarrow & \tilde{\Delta}^+_\alpha v^{++}
-(1-\partial^{\;--}_wv^{++})\tilde\omega^{+3}_\alpha =0 \label{5.26.1a} \\
{\cal D}^{\;+}_\alpha z^0=0 &\Rightarrow & \tilde{\Delta}^+_\alpha t+
\tilde\omega^{+3}_\alpha\partial^{--}_wt-\tilde\omega^{+-+}_\alpha t=0
\label{5.26.1b} \\
{\cal D}^+_\alpha w^{+i}_A &=& 0\;, \;\;\;
{\cal D}^+_\alpha w^{-i}_A = 0\;,\;\;\;
{\cal D}^+_\alpha x^{+\mu}_A = 0 \label{5.26.1c}
\eea
where ${\cal D}^+_\alpha$ and $\tilde{\Delta}^{+}_\alpha$ were defined in eqs.
\p{5.22.3} - \p{5.22.5}. Note that
first of eqs. \p{5.26.1c} is automatically satisfied as a consequence of
 \p{5.26.1a} (the second one is satisfied trivially). The
definition \p{5.26.1a} - \p{5.26.1c} admits the pregauge group with analytic
parameters
\bea
\delta  z^{++}_A=\lambda^{++}\;, \;\;\; \delta z^0_A=\lambda \; z^0_A\;,\;\;\;
\delta w^{-i}_A=0\;, \;\;\; \delta w^{+i}_A=-\lambda^{++}w^{-i}_A\;,\;\;\;
\delta x^{+\mu}_A=\lambda^{+\mu}
\label{5.26.2}
\eea
\be
{\cal D}^+_\alpha\lambda^{++}={\cal D}^+_\alpha\lambda=
{\cal D}^+_\alpha\lambda^{+\mu}=0\;,
\label{5.26.3}
\ee
while the remaining coordinates transform with general
(though $z$-independent)
parameters $\lambda^{--},\;\lambda^{-\mu}$
\bea
\delta  z^{--}_A=\lambda^{--}+2\lambda  z^{--}_A, \;\;\; \delta
x^{-\mu}_A=\lambda^{-\mu}\;.\label{5.27.1}
\eea
{}From eqs. \p{5.25.1}, \p{5.25.2}, \p{5.26.2}, \p{5.27.1} one finds the
transformations of bridges
\bea
\delta v^{++} &=& \lambda^{++}-\tau^{++}\;, \;\;\;\;
\delta v^{--}=2\tau^{+-}v^{--}-\tau^{--}+t^{-2}\lambda^{--} \nn \\
\delta v^{+\mu} &=& \lambda^{+\mu}- w^{+i}\tau^{\mu i}(x)-x^{\mu i}\tau^{++}
w^{-i}\;,\;\;\;\;
\delta v^{-\mu} = \lambda^{-\mu}- w^{-i}\tau^{\mu i}(x)\;.\nn
\eea

Finally, we should pass to the analytic $Sp(n)$ frame by rotating all
tangent space $Sp(n)$ indices by an appropriate matrix bridge. In this new
frame the tangent space group $Sp(n)$ is realized by gauge transformations
with analytic
parameters. Actually, this procedure is the same as in the HK case \cite{a2},
so we do not give here details of it.

Performing the change of variables \p{5.25.1}, \p{5.25.2} in the expressions
\p{5.10.1}, \p{5.10.2} and passing to the analytic $Sp(n)$ frame,
we find the
$\lambda$-world form of harmonic and $z$-derivatives
\bea
{\cal D}^{\;++} &=& {\cal D}^{\;++}_\lambda +
\phi^{++}(z^0_A\;{\partial\over\partial z^0_A})-
[H^{+4}-(z^{++}_A)^2]\;{\partial\over\partial z^{++}_A}\nn \\
&&+[2\phi^{++} z^{--}_A+(z^0_A)^2+\phi]\;{\partial\over\partial z^{--}_A}
+z^{++}_A\;(z^0_A{\partial\over\partial z^0_A}-{\cal D}^{\;0})
\label{5.27.3} \\
{\cal D}^{\;++}_\lambda &\equiv & \Delta^{\;++} +\omega^{++}=
\partial^{\;++}_A+H^{+3\mu}\;\partial^{\;-}_\mu +
H^{++-\mu}\;\partial^{\;+}_\mu + H^{+4}\;\partial^{\;--}_A+\omega^{++}
\label{5.27.4}\\
{\cal D}^{\;0} &=& \partial^{\;0}_A +(x^{+\mu}_A \partial^{\;-}_\mu-
x^{-\mu}_A\partial^{\;+}_\mu+2\;z^{++}_A{\partial\over\partial
z^{++}_A}-2\;z^{--}_A{\partial\over\partial z^{--}_A})
\label{5.27.5} \\
{\cal D}^{\;--} &=& (z^0_A)^{-2}\;\{\psi\;{\cal D}^{\;--}_\lambda +
[(z^0_A)^{2}-\psi]{\partial\over\partial z^{++}_A}+
(\phi^{--}+z^{--}_A)\;z^0_A{\partial\over\partial z^0_A} \nn \\
&& + [2\phi^{--}z^{--}_A+H^{-4}+(z^{--}_A)^2]\;
{\partial\over\partial z^{--}_A}\}
\label{5.28.1} \\
{\cal D}^{\;--}_\lambda &\equiv& \Delta^{--} +\omega^{--}=
\partial^{\;--}_A+H^{--+\mu}\;\partial^{\;-}_\mu +
H^{-3\mu}\;\partial^{\;+}_\mu +\omega^{--}
\label{5.28.2} \\
Z^{\;++} &=& (z^0_A)^2{\partial\over\partial z^{--}_A} \label{5.27.2} \\
Z^{\;--}&=&{\cal D}^{\;--}-(z^0_A)^{2}{\partial\over\partial z^{++}_A}
\label{5.28.3}\\
Z^0 &=& z^0_A{\partial\over\partial z^0_A}\;.\label{5.28.3a}
\eea
Here $\partial^{\;+}_{\mu},\; \partial^{\;-}_{\mu}$ stand for partial
derivatives with respect to the analytic basis coordinates
$x^{-\mu}_{A},\;x^{+\mu}_{A}$, the objects $\omega^{++},\;
\omega^{--}$ are the harmonic $Sp(n)$ connections arising as a
result of passing to the analytic frame, and the remaining quantities
(vielbeins) are defined by
\bea
H^{+3\mu} &=& \partial^{++}v^{+\mu}+v^{++}x^{+\mu}_A
\nn \\
H^{++-\mu} &=& \partial^{++}v^{-\mu} -v^{+\mu}+x^{+\mu}_A-v^{++}x^{-\mu}_A
\nn \\
H^{+4} &=& -[\partial^{++}v^{++}+(v^{++})^2]
\nn \\
\phi^{++} &=& \partial^{++}\ln t-v^{++}\nn \\
\phi &=& t^2\;(\partial^{++}v^{--}-1)\label{5.28.4e}
\eea
\bea
\psi &=& t^2\;(1-\partial^{--}v^{++})\nn \\
H^{--\pm\mu}
 &=& {1\over 1-\partial^{--}v^{++}}\;\partial^{\;--}x^{\pm\mu}_A =
\Delta^{--} x^{\pm\mu}_A \nn \\
\Delta^{--} &=& {1\over 1-\partial^{--}v^{++}}\;\partial^{\;--}
\nn \\
\phi^{--} &=& t^2\;(\partial^{\;--}\ln t-v^{--})\nn \\
H^{-4} &=& t^4\;[\partial^{--}v^{--}+(v^{--})^2]\;.
\label{5.28.5e}
\eea
The vielbeins \p{5.28.4e} - \p{5.28.5e} transform as follows
\bea
\delta H^{+3\mu} &=& \Delta^{++}\lambda^{+\mu}+\lambda^{++}x^{+\mu}_A
\nn \\
\delta H^{++-\mu} &=& \Delta^{++}\lambda^{-\mu}-\lambda^{++}x^{-\mu}_A
\nn \\
\delta H^{+4} &=& -\Delta^{++}\lambda^{++} \nn \\
\delta \phi^{++} &=& \Delta^{++}\lambda-\lambda^{++} \nn \\
\delta \phi &=& \Delta^{++}\lambda^{--}-2\;\phi^{++}
\lambda^{--}+2\;\lambda\phi
 \label{5.29.1e}
\eea
\bea
\delta\psi &=& (2\lambda-\Delta^{--}\lambda^{++})\;\psi \nn \\
\delta H^{--+\mu} &=& \Delta^{--}\lambda^{+\mu}+(\Delta^{--}\lambda^{++})\;
H^{--+\mu} \nn \\
\delta H^{-3\mu} &=& \Delta^{--}\lambda^{-\mu}+(\Delta^{--}\lambda^{++})\;
H^{-3\mu} \nn \\
\delta \phi^{--} &=& 2\lambda\;\phi^{--}-\lambda^{--}+(\Delta^{--}\lambda)\;
\psi  \nn \\
\delta H^{-4} &=& 4\lambda\;
H^{-4}-2\;\phi^{--}\lambda^{--}-(\Delta^{--}
\lambda^{--})\;\psi\;.\label{5.29.2e}
\eea
At this step it is convenient to choose the following gauges
\bea
H^{++-\mu}=x^{+\mu}_A \;\;\Rightarrow \;\;
\partial^{++}v^{-\mu} -v^{+\mu}-v^{++}x^{-\mu}_A=0,\;\;
\lambda^{+\mu}= \Delta^{++}\lambda^{-\mu}-\lambda^{++}x^{-\mu}_A
\label{5.30.1}
\eea
\bea
\phi^{--}=0 \;\;\Rightarrow \;\;v^{--}=\partial^{--}\ln t,\;\;
\lambda^{--}=\psi\Delta^{--}\lambda\;. \label{5.30.2}
\eea
In the gauge \p{5.30.2} the expression \p{5.28.1} and the last ones of the
expressions \p{5.28.4e} and \p{5.28.5e}
are somewhat simplified
\bea
{\cal D}^{\;--}=(z^0_A)^{-2}\;\{\psi{\cal D}^{--}_\lambda +
[(z^0_A)^{2}-\psi]{\partial\over\partial z^{++}_A}+
z^{--}_A(z^0_A{\partial\over\partial z^0_A})+
[H^{-4}+(z^{--}_A)^2]{\partial\over\partial z^{--}_A}\}
\label{5.28.1'}
\eea
\bea
\phi=-\psi\;(1-\Delta^{--}\phi^{++})\label{5.28.4e'}
\eea
\bea
H^{-4}=t^3\;(\partial^{--})^2t\;. \label{5.28.5e'}
\eea

The specific feature of the case under consideration compared to the HK
one is the presence of new
harmonic vielbeins $\phi^{++}$ and $\phi^{--}$ which multiply
the  $Sp(1)$ weight operator $z^0_A\partial/\partial z^0_A$
in ${\cal D}^{++}$ and
${\cal D}^{--}$. When the latter are applied to the functions
having
$p=q$ and bearing no
dependence on $z^{++}_A,\;z^{--}_A$, the new vielbeins can be interpreted
as some harmonic $Sp(1)$ connections. For instance,
on such functions ${\cal D}^{\;++}$ takes the following form
\be
{\cal D}^{\;++}={\cal D}^{\;++}_\lambda+ \phi^{++}{\cal D}^{\;0}
\label{5.30'.1}
\ee
and so it preserves the analytic subspace $\{x^{+\mu},w^{\pm i}_A\}$ (as
will be clear soon, $\phi^{++}$ and the vielbeins entering
${\cal D}^{\;++}_\lambda$ are analytic).

It is instructive to give here how the $z$-independent parts of the derivatives
${\cal D}^{\;++}$ and ${\cal D}^{\;--}$ look in the original $\tau$-basis
$\{x^{\mu i},w^{\pm i}\}$:
\bea
\Delta^{--}={1\over 1-\partial^{--}_wv^{++}}\;\partial^{\;--}_w\;,\;\;\;\;
\Delta^{++}=\partial^{--}_w+v^{++}D^{\;0}\Vert \;,
\label{5.30'.2}
\eea
where the slash means the part of $D^{\;0}$ containig no $z$-derivatives.
They are
transformed under the $\lambda$-world gauge group in the following way
(once again, when act on the
functions independent of $z^{++}_A,z^{--}_A$ and possessing $p=q$)
\bea
\delta \Delta^{++}=\lambda^{++}\;D^{\;0}\Vert,\;\;\;\;
\delta \Delta^{--}=(\Delta^{--}\lambda^{++})\;\Delta^{--}
-\Delta^{--}\lambda \;D^{\;0}\Vert\;.
\label{5.30'.3}
\eea
Note that the role of vielbein $\psi$ in eqs. \p{5.28.1} or \p{5.28.1'}
consists in converting the $Sp(1)$ weight transformation
with the nonanalytic parameter $\Delta^{++}_\lambda\lambda^{++}$ into a
transformation with the analytic parameter $\lambda$.

Finally, we quote here the general form of ${\cal D}^+_\alpha$ in the
 $\lambda$-world. Taking into account the defining relations \p{5.26.1a} -
\p{5.26.1c} and the property that this derivative does not contain $Sp(n)$
connection after passing to the $\lambda$-frame (cf.
the similar situation in the HK case \cite{a2}), it is given
by
\bea
{\cal D}^+_{A\alpha}=E^\mu_\alpha\partial^+_\mu+\rho^-_\alpha
 {\partial\over\partial z^{--}_A}\equiv \Delta^+_\alpha +\rho^-_\alpha
 {\partial\over\partial z^{--}_A} = (z^0_A)\;(\tilde{\Delta}^{+}_{\alpha} +
\tilde{\rho}^{-}_{\alpha}{\partial\over\partial z^{--}_{A}})\;.
\label{5.31.1}
\eea

The transformation properties of this important
object and the equations for the vielbeins $E^\mu_\alpha$ and $\rho^-_\alpha$
following from the quaternionic geometry constraints \p{5.20.3a} -
\p{5.20.3d}, \p{5.21.1} will be discussed in the next section.

\setcounter{equation}{0}
\section{$\lambda$-world geometry: consequences of constraints}

Our ultimate purpose will be to draw all the
consequences the defining
constraints \p{5.20.3a}-\p{5.20.3d}, \p{5.21.1} entail for
various vielbeins and connections in the $\lambda$-world and to express (in
the next Section) the
latter in
terms of a few unconstrained prepotentials. As we will see below,
these prepotentials coincide with certain vielbeins multiplying the
derivatives with respect to $z$-coordinates.
This automatic appearance of the prepotentials as the
objects having a clear geometrical meaning is the actual advantage of dealing
with the $z$-extension of the quaternionic manifolds.

Recall that
the HK geometry prepotentials \cite{a2} can be deduced in two equivalent ways:
either as some new objects in the process of solving the
HK constraints or as the vielbeins associated with an extension of the
original HK manifold by a kind of ``central charge'' coordinates.
In the quaternionic case just the second, most geometric approach proves
to be adequate,
because, as we saw in previous Sections, it is {\it necessary} to extend the
initial manifold by $z$-coordinates for giving the CR-structure interpretation
to the quaternionic geometry constraints.

\subsection{Derivative ${\cal D}^+_{A\alpha}$}

We begin with studying the covariant derivative ${\cal D}^+_{A\alpha}$
defined in eq. \p{5.31.1}. The vielbeins $E_\alpha^\mu$ and
$\rho^-_\alpha$ transform under the analytic tangent space
transformations and diffeomorphisms \p{5.27.1} according to
\bea
\delta E_\alpha^\mu &=& \lambda_\alpha{}^\beta \;E_\beta^\mu
+E_\alpha^\nu\;\partial^{\;+}_\nu\lambda^{-\mu} \label{5.32.1} \\
\delta\rho^-_\alpha &=& 2\;\lambda\;\rho^-_\alpha+
E_\alpha^\mu\;\partial^{\;+}_\mu\lambda^{--}+\lambda_\alpha{}^\beta \;
\rho^-_\beta\;.
\label{5.32.2}
\eea
One should remember that ${\cal D}^+_{A\alpha}$ has the $Sp(1)$ weght
$1$ and hence it is linear in the coordinate $z^0_A$. The
$z$-independent vielbeins ${\tilde E}_\alpha^\mu=
(z^0_A)^{-1}\;E_\alpha^\mu$ and ${\tilde\rho}^-_\alpha=
(z^0_A)^{-1}\;\rho^-_\alpha$ transform under the $Sp(1)$ $\lambda$-
transformations as
\be
\delta {\tilde E}_\alpha^\mu=-\lambda\;{\tilde E}_\alpha^\mu\;, \;\;
\delta {\tilde\rho}^-_\alpha=\lambda\;{\tilde\rho}^-_\alpha. \label{5.32.3}
\ee

The basic constraint \p{5.20.3a}
\be
[{\cal D}^{\;+}_{A\alpha},\;{\cal D}^{\;+}_{A\beta}]=-
2\;\Omega_{\alpha\beta}R(z^0_A)^{2}
{\partial\over\partial z^{--}} \nn
\ee
implies the following relations
\bea
E^\mu_{[\alpha}\;\partial^{\;+}_{A\mu}E^\rho_{\beta]} &=& 0 \label{5.33.1} \\
\Delta^+_{[\alpha}\;\rho^-_{\beta]} = -
\Omega_{\alpha\beta}\;R\;(z^0_A)^{2}\;,
 \;\;\;\; &{\rm or}& \;\tilde\Delta^+_{[\alpha}\rho^-_{\beta]} = -
\Omega_{\alpha\beta}\;R\;.
\label{5.33.2}
\eea
Eq. \p{5.33.1} is the same as in the HK case while \p{5.33.2} is new.

The next important constraint is
\be
[{\cal D}^{\;+}_{A\alpha},\;{\cal D}^{\;++}]=0\;. \label{5.33.3}
\ee
It implies the analyticity of the vielbeins and connections entering
into ${\cal D}^{++}$
\bea
\Delta^{\;+}_{\alpha}\;H^{+3\mu} = \Delta^{\;+}_{\alpha}\;H^{+4} =
\Delta^{\;+}_{\alpha}\;\omega^{++} =
\Delta^{\;+}_{\alpha}\;\phi^{++} = 0 \label{5.33.4}
\eea
and also results in the relation
\be
{\cal D}^{\;++}_\lambda E_\alpha^\mu +\phi^{++}\;E_\alpha^\mu=0\;,
\label{5.33.5}
\ee
or, equivalently,
\be
E_{(\alpha}^\mu \;\Delta^{++} E^{-1}_{\mu\beta)} =
\omega^{++}_{\alpha\beta}
 \label{5.33.5a}
\ee
\be
E_{[\alpha}^\mu \;\Delta^{++} E^{-1}_{\mu\beta]}=
\Omega_{\beta\alpha}\;\phi^{++} \;,\label{5.33.5b}
\ee
and in the following one
\be
{\cal D}^{\;++}_\lambda\rho^-_\alpha -\phi^{++}\rho^-_\alpha
 -\Delta^{\;+}_\alpha\phi=0\;.  \label{5.34.1}
\ee
Note the appearance of a non-zero right-hand side in eq. \p{5.33.5b}
in contrast to an analogous equation in the HK case. The constraints
\p{5.33.1}, \p{5.33.2}, \p{5.33.5b}, \p{5.34.1} will be solved later on,
after defining the derivative ${\cal D}^{\;-}_{A\alpha}$.

\subsection{Consequences of the algebra of harmonic derivatives}

Like in the HK case, the vielbeins and connections entering the harmonic
derivative ${\cal D}^{--}$ can be related to the analytic vielbeins
\p{5.33.4} by the equation
\be
[{\cal D}^{\;++},{\cal D}^{\;--}]={\cal D}^{\;0} \;.\label{5.34.2}
\ee
Comparing the coefficients of the same derivatives in both
sides of \p{5.34.2}, one finds
\be
\Delta^{--}H^{+4}\;=\;\Delta^{++}\ln\psi -2\phi^{++} \label{5.34.3a}
\ee
\bea
\Delta^{++}H^{-4}-4\;\phi^{++}H^{-4}-\psi\;\Delta^{--}\phi &=& 0
\label{5.34.3b} \\
(\Delta^{++}+\Delta^{--}H^{+4})\;H^{--+\mu}-
\Delta^{--}H^{+3\mu} &=& x^{+\mu}_A \nn \\
(\Delta^{++}+\Delta^{--}H^{+4})\;H^{-3\mu}-H^{--+\mu} &=& x^{-\mu}_A \nn \\
(\Delta^{++}+\Delta^{--}H^{+4})\;\omega^{--} -\Delta^{--}\omega^{++}+
[\;\omega^{++},\omega^{--}\;] &=& 0\;. \label{5.35.1c}
\eea
Using eqs. \p{5.34.3a} - \p{5.35.1c} one may, at least iteratively,
express $H^{--+\mu},\;H^{-3\mu},\;\omega^{--},\;\psi,\; H^{-4}$ in terms of
$H^{+3\mu},\; H^{+4},\; \phi^{++}$ and $\omega^{++}$.
Note the useful
relation
\be
[\;{\cal D}^{\;++}_\lambda,\; {\cal D}^{\;--}_\lambda\;] =
{\cal D}^{\;0}_\lambda -(\Delta^{--} H^{+4})\;{\cal D}^{\;--}_\lambda \;,
\label{5.35.2}
\ee
which follows from eqs. \p{5.35.1c}.

\subsection{Derivative ${\cal D}^-_\alpha$}

We define the $\lambda$-world derivative  ${\cal D}^-_\alpha$ on the
pattern already employed in the $\tau$-world (eq. \p{5.19.3})
\bea
{\cal D}^-_\alpha=[{\cal D}^{--},{\cal D}^+_\alpha] &=&
(z^0_A)^{-2}\;\{(E^{-2\mu}_\alpha +z^{--}_A
E^\mu_\alpha)\;\partial^{\;+}_\mu +
E^{-+\mu}_\alpha \partial^{\;-}_\mu +E^+_\alpha\;(\partial^{--}_A-
{\partial\over\partial z^{++}_A})  \nn \\
&& + (E^{-3}_\alpha-z^{--}_A\rho^-_\alpha)\;{\partial\over\partial z^{--}_A}
-\rho^-_\alpha\;(z^0_A{\partial\over\partial z^0_A}) +\omega^-_\alpha\}\;,
\label{5.35.3}
\eea
where
\bea
E^{-2\mu}_\alpha &=& \psi\;{\cal D}^{\;--}_\lambda E^\mu_\alpha -
\Delta^+_\alpha\psi\; H^{-3\mu}-\psi\;\Delta^+_\alpha  H^{-3\mu}
\equiv z^0_A\;{\tilde E}^{-2\mu}_\alpha \label{5.35.4a} \\
E^{-+\mu}_\alpha &=& -(\Delta^+_\alpha\psi\; H^{--+\mu}+
\psi\;\Delta^+_\alpha  H^{--+\mu}) \equiv e_\alpha{}^\mu=
 z^0_A\;\tilde{e}_\alpha{}^\mu \label{5.35.4b} \\
E^+_\alpha &=& - \Delta^+_\alpha\psi \equiv -{\cal L}^+_\alpha= -
 z^0_A\;{\tilde{\cal L}}^+_\alpha \label{5.35.4c} \\
E^{-3}_\alpha &=& \psi\;{\cal D}^{--}_\lambda\rho^-_\alpha-
\Delta^+_\alpha H^{-4}
\equiv z^0_A\;{\tilde E}^{-3}_\alpha  \label{5.35.4d} \\
\omega^-_\alpha &=& -\psi\;(\Delta^+_\alpha \ln\psi \;\omega^{--}+
\Delta^+_\alpha \omega^{--}) \equiv  z^0_A\;{\tilde\omega}^-_\alpha\;.
\label{5.35.4e}
\eea

A number of important constraints on these objects follows from
considering the commutator
\bea
[{\cal D}^{\;+}_\alpha,\;{\cal D}^{\;-}_\beta]_{\rho}{}^{\rho'}=
\delta_{\rho}{}^{\rho'} \;\Omega_{\alpha\beta}\;R\;Z^{\;0}
+(R^{+-}_{\alpha\beta})_\rho{}^{\rho'}
\label{5.36.1}
\eea
The vanishing of the torsion components $T^{+-(+\mu)}_{\alpha\beta}$ and
$T^{+-(+2)}_{\alpha\beta}$ in the r.h.s. of \p{5.36.1} ( the
coefficients of $\partial^{\;-}_\mu$ and
 $\partial^{\;--}$) implies, respectively,
\bea
\Delta^+_\alpha e^\mu_\beta = 0\;,\;\;\;\;\;
\Delta^+_\alpha {\cal L}^+_\beta &=& 0\;,  \label{5.36.3}
\eea
i.e., $e^\mu_\beta$ and ${\cal L}^+_\beta$ are analytic. Equating the
coefficients of $Z^{\;0}$ in both sides of \p{5.36.1} and taking account of
eq.\p{5.33.2}, one finds
\be
\Delta^+_{(\alpha}\rho^-_{\beta)}=0\;. \label{5.36.4}
\ee
Also, it is easy to evaluate the $Sp(n)$ curvature $R^{+-}_{\alpha\beta}$
\bea
(R^{+-}_{\alpha\beta})_\rho{}^{\rho'}=
(\tilde{\Delta}^+_\alpha \tilde\omega^-_\beta)_\rho{}^{\rho'}=
\{\psi^{-1}\;\tilde{\cal L}^+_\alpha \;\tilde\omega^-_\beta+
\psi\tilde\Delta^+_\alpha(\psi^{-1}\tilde\omega^-_\beta)\}_\rho{}^{\rho'}\;,
\label{5.36.5}
\eea
where, as before, tilde means that the relevant quantity does not
contain the dependence on $z^0_A$. Using the explicit expression for
$\tilde\omega^-_\beta$ \p{5.35.4e} one may easily be convinced that
$R^{+-}_{\alpha\beta}$ is symmetric in the indices $\alpha,\; \beta$
\be
R^{+-}_{[\alpha\beta]}=\tilde\Delta^+_{[\alpha}\tilde\omega^-_{\beta]}=0\;.
\label{5.36.6}
\ee

There is also a number of additional relations following from
eq.\p{5.36.1}, but all these can be checked to be consequences of a
few essential ones some of which have been already given and others
will be presented below.

\subsection{Vielbein $E^\mu_\alpha$}

Using eq. \p{5.35.4b}, one may express the vielbein $E^\mu_\alpha$
defined in eq. \p{5.31.1} through the analytic quantities
$e^\mu_\alpha$, ${\cal L}^+_\alpha$ and the harmonic vielbeins $\psi$
and $H^{--+\mu}$:
\bea
E^\mu_\alpha=\psi^{-1}A_\alpha{}^\nu(\partial H^{-1})_\nu{}^\mu \;.
\label{5.37.1}
\eea
Here
\bea
A_\alpha{}^\nu &\equiv & e_\alpha{}^\nu-{\cal L}^+_\alpha H^{--+\nu}
\label{5.37.2} \\
(\partial H)_\nu{}^\mu &\equiv &
\partial^+_\nu H^{--+\mu}. \label{5.37.3}
\eea
After some simple algebra $E^\mu_\alpha$ can be concisely written as
\be
E^\mu_\alpha={1\over \psi(1-{\cal L}H)}\;
e_\alpha{}^\rho(\partial {\hat H}^{-1})_\rho{}^\mu \;, \label{5.37.4}
\ee
where
\bea
{\cal L}H \equiv {\cal L}^+_\mu H^{--+\mu}\;, \;\;\;\;
 {\cal L}^+_\mu  \equiv  e_\mu{}^\alpha {\cal L}^+_\alpha\;, \;\;\;\;
 e_\mu{}^\beta  e_\alpha{}^\mu= \delta^\beta_\alpha
\label{5.37.5}
\eea
\be
{\hat H}^{--+\rho} \equiv {1\over 1-{\cal L}H}\;H^{--+\rho}.\label{5.37.6}
\ee
It is a simple exercise to check that the constraint \p{5.33.1} is
identically satisfied by the expression \p{5.37.4}.

Before going further, let us quote the transformation properties of
the important quantities $e^\mu_\alpha$, $e_\mu{}^\alpha$, ${\cal L}^+_\mu$
introduced in subsec. 6.7 and 6.8
\bea
\delta e_\alpha{}^\mu &=& \lambda_\alpha{}^\beta\; e_\beta{}^\mu
 +2\;\lambda\; e_\alpha{}^\mu +e_\alpha{}^\rho\;
\hat\partial_\rho^{\;-}\lambda^{+\mu}
\nn \\
\delta e_\nu{}^\alpha &=& -e_\nu{}^\beta \;\lambda_\beta{}^\alpha-
2\;\lambda \;e_\nu{}^\alpha- \hat\partial_\nu^{\;-}
\lambda^{+\mu}\;e_\mu{}^\alpha \nn\\
\delta{\cal L}^+_\mu &=& -\hat\partial_\mu^{\;-}\lambda^{+\nu}\;{\cal L}^+_\nu
-\hat\partial_\mu^{\;-}\lambda^{++}\;, \label{5.38.1}
\eea
where
\bea
\hat\partial_\mu^{\;-} &\equiv &
\partial_\mu^{\;-} +{\cal L}^+_\mu\;\partial^{\;--}
\label{5.38.2} \\
\left[ \hat\partial_\mu^{\;-},\;\hat\partial_\nu^{\;-} \right] &=&
2\; \hat\partial_{\;[\mu^-}{\cal L}^+_{\;\nu]}\;\partial^{\;--}
\equiv 2\;T_{\mu\nu}\;\partial^{\;--} \label{5.38.3}
\eea
\bea
\delta\hat\partial_\mu^{\;-} &=&
- \hat\partial_\mu^{\;-}\lambda^{+\rho}\;\hat\partial_\rho^{\;-} \nn \\
\delta T_{\mu\nu} &=& -\hat\partial_\mu^{\;-}\lambda^{+\rho}\;T_{\rho\nu}+
\hat\partial_\nu^{\;-}\lambda^{+\rho}\;T_{\rho\mu}-
T_{\mu\nu}\;(\partial^{\;--}\lambda^{++}+
{\cal L}^+_\rho\;\partial^{\;--}\lambda^{+\rho})\;. \label{5.38.4} \eea

One immediately checks that $E_\alpha{}^\mu$ given in \p{5.37.4}
have the necessary transformation properties \p{5.32.1}.

\setcounter{equation}{0}
\section{$\lambda$-world geometry: quaternionic potentials}

In this Section we continue studying the quaternionic geometry in the
$\lambda$-world representation. Here we at last reach the place
where the quaternionic geometry potentials appear.

\subsection{Expressing $\omega^{++}_{\alpha\beta}$, $\phi^{++}$ and
$H^{+3\nu}$}

A new set of important relations arises from considering the
commutator
\be
[{\cal D}^{++}, {\cal D}^-_\alpha]={\cal D}^+_\alpha \label{5.39.1}
\ee
which is trivially satisfied in the $\tau$-world but leads to
nontrivial restrictions while treating the $\lambda$-world geometry.
The most informative
relations follow from equating the coefficients of $\partial^{\;-}_\mu$
and $\partial^{\;--}$ in both sides of \p{5.39.1}
\be
{\cal D}^{\;++}_\lambda e_\alpha{}^\mu -\phi^{++}\;e_\alpha{}^\mu-
e_\alpha{}^\nu\;\hat\partial_\nu^{\;-}H^{+3\mu}-{\cal L}^+_\alpha
\;x^{+\mu}_A=0,
\label{5.39.2}
\ee
\be
{\cal D}^{\;++}_\lambda {\cal L}^+_\alpha-\phi^{++}\;{\cal L}^+_\alpha-
e_\alpha{}^\mu \;\hat\partial_\mu^{\;-}H^{+4}=0 \label{5.39.3}
\ee
All the other equations associated with \p{5.39.1} can be shown to be
satisfied in virtue of eqs. \p{5.39.2}, \p{5.39.3} and those deduced
previously.

Let us turn to extracting the consequences of eqs. \p{5.39.2}, \p{5.39.3}.

As a first corollary we note that the previously written constraint
\p{5.33.5} is satisfied identically if one substitutes for $E_\alpha{}^\mu$
its expression \p{5.37.4} and takes into account eqs. \p{5.39.2}, \p{5.39.3}
and \p{5.34.3a}. So we may forget about eq. \p{5.33.5}. At this step we may
also express the connections $\omega^{++}_{\alpha\gamma}$ and $\phi^{++}$
in terms of the analytic vielbeins
\bea
\omega^{++}_{\alpha\gamma} &=& e_{(\alpha}{}^\mu\;\Delta^{++}e_{\mu\gamma)}
+e_{(\alpha}{}^\nu\;\hat\partial_\nu^{\;-}H^{+3\mu}\;e_{\mu\gamma)}
+{\cal L}^+_{(\alpha}\;x^{+\mu}_A\;e_{\mu\gamma)} \label{5.40.1} \\
\phi^{++} &=& {1\over 2n}\;(e_\mu{}^\alpha\Delta^{++}e_{\alpha}{}^\mu
-\hat\partial_\nu^{\;-}H^{+3\nu}-{\cal L}^+_\mu x^{+\mu}_A)\;.\label{5.40.2}
\eea
Further, let us  rewrite \p{5.39.2}, \p{5.39.3} as the equations for
$e_\mu{}^\alpha$, ${\cal L}^+_\mu$
\bea
{\cal D}^{\;++}_\lambda e_\mu{}^\alpha+\phi^{++}\;e_\mu{}^\alpha+
e_\nu{}^\alpha\;\hat\partial_\mu^{\;-}H^{+3\nu}+
e_\nu{}^\alpha \;x^{+\nu}_A\;{\cal L}^+_\mu=0 \label{5.39.2'}
\eea
\bea
\Delta^{++}{\cal L}^+_\mu+{\cal L}^+_\nu\;\hat\partial_\mu^{\;-}H^{+3\nu}
+{\cal L}^+_\mu\;({\cal L}^+_\nu x^{+\nu}_A)-\hat\partial_\mu^{\;-}H^{+4}=0\;.
\label{5.39.3'}
\eea
Then, introducing
\be
{\cal L}^{+4}\equiv H^{+4}-{\cal L}^+_\nu \;H^{+3\nu} \label{5.40.3a}
\ee
\bea
\delta{\cal L}^{+4}=-\partial^{++}\lambda^{++}
-{\cal L}^{+4}(\partial^{--}\lambda^{++})
-\lambda^{++}({\cal L}^+_\nu x^{+\nu}_A)
-{\cal L}^+_\nu \partial^{++}\lambda^{+\nu}-
({\cal L}^+_\nu \partial^{--}\lambda^{+\nu}){\cal L}^{+4}\;, \label{5.40.3}
\eea
one may express the vielbein $ H^{+3\nu}$ from eq. \p{5.39.3'} as a
function of ${\cal L}^{+4}$ and ${\cal L}^+_\mu$
\bea
 H^{+3\nu}=-{1\over 2}\;T^{\nu\mu}\;[\;\hat\partial_\mu^{\;-}{\cal L}^{+4}-
\partial^{\;++}{\cal L}^+_\mu
 -{\cal L}^{+4}\;\partial^{\;--}{\cal L}^+_\mu
-{\cal L}^+_\mu\;({\cal L}^+_\nu x^{+\nu}_A)\;]\;, \label{5.41.1}
\eea
where $T^{\nu\mu}$ is inverse to $T_{\mu\nu}$ defined in
eq.\p{5.38.3},  $T^{\nu\mu}T_{\mu\rho}=\delta^\nu_\rho $. \\

Thus we have succeeded in expressing the harmonic vielbein  $ H^{+3\nu}$
and, in virtue of eqs. \p{5.35.1c}, the vielbeins entering
into ${\cal D}^{--}_\lambda$, through two {\it analytic} and otherwise
{\it unconstrained} objects, ${\cal L}^+_\mu(x^+_A,w^{\pm i})$ and
${\cal L}^{+4}(x^+_A,w^{\pm i})$. These objects, just as in the HK
case, are {\it fundamental} objects of the quaternionic geometry, its
{\it potentials}. As we see, the real advantage of considering the
bi-harmonic extension of the quaternionic manifold is that the
prepotentials naturally appear as the vielbeins associated with
the partial derivative $(\partial^{\;--}-\partial/\partial z^{++})$ in
the covariant derivatives ${\cal D}^{\;++}$ \p{5.27.3}  and
${\cal D}^{\;-}_\alpha$ \p{5.35.3}.

\subsection{Fixing analytic $Sp(1)$ weight gauge freedom}

Below we will demonstrate that all the
geometric
objects encountered earlier in this and previous Sections can be expressed
through
${\cal L}^+_\mu(x^+_A,w^{\pm i})$ and ${\cal L}^{+4}(x^+_A,w^{\pm i})$
after fixing a gauge with respect to the analytic
$Sp(1)$ weight transformation (with the parameter $\lambda(x^+_A,w^{\pm i})$).

To see that this gauge freedom can be completely fixed, let us pass to
${\tilde e}_\alpha{}^\mu$ and ${\tilde A}_\alpha{}^\mu$ (with $z^0_A$
factored out) and define
\be
{\tilde e} \equiv \det{\tilde e}_\alpha{}^\mu,\;\;\;\;{\tilde A} \equiv
\det {\tilde A}_\alpha{}^\mu
 = (1-{\cal L}^+_\mu H^{--+\mu})\;\tilde{e}
\label{5.42.3}
\ee
\bea
\delta{\tilde e} = (\hat\partial_\mu^{\;-}
\lambda^{+\mu}+2n\lambda)\;{\tilde e}\;,\;\;\;\;
\delta{\tilde A} = (\partial_\mu^{\;-}\lambda^{+\mu}+
H^{--+\rho}\;\partial_\rho^{\;-}\lambda^{++}+2n\lambda)\;
{\tilde A}\;. \label{5.42.2}
\eea
Next, one introduces
\be
B=\psi{\tilde A}, \;\;\;\;\;
\delta B =[2(n+1)\lambda+\partial_\mu^-\lambda^{+\mu}
-\partial^{--}\lambda^{++}]B\label{5.42.4}
\ee
and checks that $B$ is analytic,
\be
\Delta^+_\alpha B=0 \;.\label{5.42.5} \ee
Thus, we have two independent analytic quantities, $B$ and $\tilde{e}$,
possessing nontrivial
and different transformation laws under the analytic $Sp(1)$ weight
transformations. Hence one may completely fix the gauge with respect to these
transformations by gauging away some combination of $B$ and $\tilde{e}$.
There exists a two-parameter family of such gauges
\be
B^\gamma \;e^\alpha= 1 \;\;\;\Rightarrow  \label{5.43.1}
\ee
\bea
\lambda=
{\gamma\;\partial^{\;--}\lambda^{++}
-\alpha\;{\cal L}^+_\mu\;\partial^{\;--}\lambda^{+\mu}-
(\gamma+\alpha)\;\partial_\mu^{\;-}\lambda^{+\mu}\over
2[\gamma(n+1)+\alpha n]}\;,
 \;\;\;\;\;\;
\gamma(n+1)+\alpha n \neq 0\;. \label{5.43.2}
\eea
As we will see, the most convenient choice is
\bea
\gamma = -\alpha \;\;\;\;\Rightarrow B= {\tilde e}\;,
\;\;\;\;\;\lambda = {1\over 2}\;(\partial^{\;--}\lambda^{++}+
{\cal L}^+_\mu\;\partial^{\;--}\lambda^{+\mu})\;. \label{5.43.3}
\eea

Expressing $\psi$ from eqs. \p{5.42.3}, \p{5.42.4}
\be
\psi= {B\over{\tilde e}}{1\over 1-{\cal L}H}\;, \label{5.43.4}
\ee
one finds that in the gauge \p{5.43.3} $\psi$ is a function entirely of
the prepotentials ${\cal L}^+_\mu$ and ${\cal L}^{+4}$
\be
\psi={1\over 1-{\cal L}H}\;. \label{5.43.5}
\ee

Below we will see that even before fixing the gauge one of the
quantities $B$ and ${\tilde e}$ can be expressed through the
quaternionic potentials.

\subsection{Expressing $\rho^-_\mu$ and $e_\mu{}^\alpha$}

In this subsection we will express the vielbeins
$\rho^-_\mu$ and $e_\mu{}^\alpha$ (defined by eqs. \p{5.31.1} and
\p{5.35.4b}) in terms of the prepotentials.

To this end we need to consider the commutator
\bea
[{\cal D}^{--},\;{\cal D}^-_\alpha]=0 \label{5.44.1}
\eea
whose validity is evident while it is considered  in the $\tau$-world.
The relations relevant to our purpose follow from vanishing of the
torsion components $T^{-3(++)}_{\;\alpha}$ and
$T^{-3(+\mu)}_{\;\alpha}$ :
\bea
T^{-3(++)}_{\;\alpha}=0 &\Rightarrow &
 {\cal D}^{\;--}_\lambda {\cal L}^+_\alpha+
E_\alpha^{-2\mu}\;\partial^{\;+}_\mu\ln\psi -e_\alpha{}^{\rho}\;
\hat\partial^{\;-}_\rho\ln\psi + 2\;\rho^-_\alpha=0 \label{5.44.2a} \\
T^{-3(+\mu)}_{\;\alpha}=0  &\Rightarrow &
{\cal D}^{\;--}_\lambda e_\alpha{}^\mu-
{\cal D}^{\;--}_\lambda{\cal L}^+_\alpha\;H^{--+\mu}
+E_\alpha^{-2\rho}\;\partial^{\;+}_\rho H^{--+\mu} \nn \\
&& - e_\alpha{}^\rho\;\hat\partial^{\;-}_\rho H^{--+\mu} = 0\;, \label{5.44.2b}
\eea
where $E_\alpha^{-2\rho}$ is given by \p{5.35.4a}. All the remaining  relations
are satisfied either as cosequences of these essential ones or in virtue of
those obtained earlier.

Expressing $E_\alpha^{-2\rho}$ from eq. \p{5.44.2b} and substituting the
result
into eq. \p{5.44.2a} one gets the expression for $\rho^-_\mu\equiv
e_\mu{}^\alpha \rho^-_\alpha = \tilde{e}_\mu{}^\alpha \tilde{\rho}^-_\alpha$:
\bea
\rho^-_\mu=-{1\over 2}\;(\;\hat\partial^{\;-}_\mu\ln{\tilde{e}\over B}+
\partial^{--}
{\cal L}^+_\mu+2{\hat H}^{--+\nu}T_{\nu\mu}\;)\;.   \label{5.45.1}
\eea
It is a simple exercise to check that $\rho^-_\mu$ possesses the needed
transformation properties
\be
\delta\rho^-_\mu=\hat\partial^{\;-}_\mu\lambda-\hat\partial^{\;-}_\mu
\lambda^{+\nu}
\;\rho^-_\nu \;.\label{5.45.2}
\ee

We leave it to the reader to prove the validity of the relation \p{5.34.1}.
One must take into account the relation \p{5.28.4e'} whence
\bea
\Delta^+_\alpha\phi = -\Delta^+_\alpha\;[\;\psi\;(1-\Delta^{--}\phi^{++})\;]
 = e_\alpha{}^\mu\;\hat\partial^{\;-}_\mu\phi^{++}-{\cal L}^+_\alpha\;, \nn
\eea
use one more reprsentation for the harmonic $Sp(1)$ connection $\phi^{++}$
following from eq.\p{5.34.3a}
\bea
\phi^{++}={1\over 2}\;(\;{\cal L}^+_\nu\partial^{--} H^{+3\nu}+{\cal L}^+_\mu
\;x^{+\mu}_\nu-\partial^{\;--} H^{+4}-\Delta^{++}\ln{\tilde{e}\over
B}\;)\label{5.45.3} \eea
and, finally, enforce the gauge \p{5.43.3}.

It remains to find the appropriate expression for $e_\mu{}^\alpha \equiv
(z_{A}^{0})^{-2}\tilde{e}_\mu{}^\alpha$. Like in the
HK case, it is more convenient to deal with the ``metric'' $h_{\mu\nu}
\equiv (z^{0}_{A})^{-2}\tilde{h}_{\mu\nu}$
\bea
h_{\mu\nu}\equiv e_\mu{}^\alpha e_{\nu\alpha}=-h_{\nu\mu}\;,\;\;\;\;
{\tilde h}_{\mu\nu}\equiv {\tilde e}_\mu{}^\alpha {\tilde e}_{\nu\alpha}\;,
\;\;\;\; \delta{\tilde h}_{\mu\nu}=-
\hat\partial^{\;-}_\mu\lambda^{+\rho}\;{\tilde h}_{\rho\mu}
-2\;\lambda\;{\tilde h}_{\mu\nu}\;.\label{5.45.5} \eea
Being rewritten for $\tilde{h}_{\mu\nu}$, eq. \p{5.39.2} takes the form
\bea
\Delta^{++} \tilde{h}_{\rho\mu}+2\;\phi^{++}\;\tilde{h}_{\rho\mu}-
2\;\hat\partial^{\;-}_{[\;\rho}H^{+3\sigma}
\tilde{h}_{\mu]\sigma}-2\;{\cal L}^+_{[\;\rho}\;
x^{+\nu}_A \tilde{h}_{\mu]\nu}=0\;.\label{5.46.1}
\eea

Let us now apply to the constraints \p{5.33.2}, \p{5.36.4} for
the vielbein $\rho^-_\mu$. Put together, they can be rewritten as
\be
e_\mu{}^\alpha\Delta^+_\alpha\rho^-_\nu
=\tilde{e}_\mu{}^\alpha\tilde{\Delta}^+_\alpha\rho^-_\nu
= -{\tilde h}_{\mu\nu}\;R
\label{5.46.2}\ee
Substituting \p{5.45.1} into \p{5.46.2} and keeping in mind the
analyticity of $\tilde{e}/B$, ${\cal L}^+_\mu$ and $T_{\mu\nu}$, one gets,
after some algebra
\be
{\tilde h}_{\mu\nu}={R}^{-1}\;\left( {\tilde e\over B} \right)\;T_{\mu\nu}\;,
\label{5.46.3}\ee
whence
\be
\tilde e=(\det T)^{-{1\over 2(n+1)}}\;B^{{n\over n+1}}\;
(R)^{{n\over n+1}}\;. \label{5.46.4}\ee

Thus, we have expressed ${\tilde h}_{\mu\nu}$ (and, hence,
$\tilde{e}_{\mu}^{\alpha}$, up to an analytic frame $Sp(n)$ rotation)
in terms of the
prepotential ${\cal L}^+_\mu$ and the analytic ``compensator'' $B$. In
the gauge \p{5.43.3}
\be
{\tilde h}_{\mu\nu}= R^{-1}\;T_{\mu\nu}\;,
\label{5.46.5}\ee
\be
\tilde e=(\det T)^{-{1\over 2}}\;R^n\;. \label{5.46.6}\ee

It remains to check the identity \p{5.46.1}. Using the explicit form
\p{5.38.3} of $T_{\mu\nu}$, it is easy to obtain
\bea
\Delta^{++} T_{\mu\nu}-2\;\hat\partial^{\;-}_{[\;\mu}H^{+3\rho}\;T_{\nu]\rho}
-2\;{\cal L}^+_{[\;\mu}\;x^{+\rho}\;T_{\nu]\rho} && \nn \\
+T_{\mu\nu}\;({\cal L}^+_\rho \;x^{+\rho}
-\partial^{\;--}H^{+4}+{\cal L}^+_\rho \;\partial^{\;--}H^{+3\rho}) &=& 0\;.
\label{5.47.1}\eea
Remembering \p{5.45.3}, one observes that the expression within the
parentheses
is just $2\phi^{++}$ in the gauge \p{5.43.3}. This proves
\p{5.46.1}.

\subsection{Full structure of ${\cal D}^{\;--}$ and
${\cal D}^{\;-}_{\alpha}$}

To finish expressing the objects of the quaternionic geometry through
analytic potentials, we have to completely specify the structure of the
derivatives ${\cal D}^{\;--}$ and ${\cal D}^{\;-}_\alpha$ by finding
appropriate expressions for the $Sp(n)$ connection
$\omega^{--}$ and
the vielbein $E^{-2\mu}_\alpha$ defined in eq. \p{5.35.4a}. To this end,
we need to express ${\cal D}^{\;--}_\lambda E^{\mu}_\alpha$. Comparing
the representation for ${\cal D}^{\;--}_\lambda E^{\mu}_\alpha$ following
from
eq. \p{5.44.2b} with the result of the explicit action of
 ${\cal D}^{\;--}_\lambda $ on the expression \p{5.37.1}, one finds
\bea
({\cal D}^{\;--}_\lambda E^{\mu}_\alpha )\; E_{\mu\beta} &=& -{1\over 2}\;
[\;\Omega_{\beta\alpha}\Delta^{--}\ln\psi -
A_\alpha{}^\sigma(\partial^{\;-}_\sigma H^{--+\rho})A_{\rho\beta} \nn \\
&&-A_\alpha{}^\mu(\partial H^{-1})_\mu{}^\sigma
\partial^{\;+}_\sigma H^{-3\nu}\;\partial^-_\nu H^{--+\rho}A_{\rho\beta} \nn \\
&& + A_\alpha{}^\mu(\partial H^{-1})_\mu{}^\sigma\;\Delta^{--}\partial^+_\sigma
H^{--+\rho}\;A_{\rho\beta}\;]\;.
 \label{5.47.2}\eea
Further, let us apply the derivative ${\cal D}^{\;++}={\cal
D}^{\;++}_\lambda +\phi^{++}\;Z^{\;0}$ (we are not interested in terms
with the $z^{\pm\pm}$ derivatives) to
$(z^0_A)^{-2}\;\psi\;({\cal D}^{--}_\lambda E\; E^{-1})_{\alpha\beta}$.
Using the commutation relation \p{5.35.2} and the constraint
\p{5.34.3a}, one gets
\bea
{\cal D}^{\;++}\;[(z^0_A)^{-2}\;\psi\;({\cal D}^{--}_\lambda E\;
E^{-1})_{\alpha\beta}]
=\Omega_{\alpha\beta}\;(z^0_A)^{-2}\;\psi\;\Delta^{--}\phi^{++}\;,
\label{5.48.1}\eea
whence it immediately follows  that
\bea
{\cal D}^{\;--}_\lambda E^{\mu}_{(\alpha}\; E_{\mu\beta)}=0
\label{5.48.2a}\eea
\bea
{\cal D}^{--}_\lambda E^{\mu}_{[\alpha}\; E_{\mu\beta]}
\sim \Omega_{\alpha\beta}.
\label{5.48.2b}\eea
{}From eq. \p{5.48.2a} we find $\omega^{--}_{\alpha\beta}$ as a function
of $E_\alpha{}^\mu$ and, hence, of the prepotentials ${\cal L}^{+4}$
and ${\cal L}^{+}_\alpha$
\be
\omega^{--}_{\alpha\beta}=E_{(\alpha}{}^\mu\;\Delta^{--}
E^{-1}_{\mu\beta)}\;, \label{5.48.3}\ee
while eqs. \p{5.48.2b} and \p{5.47.2} put together yield
\be
{\cal D}^{\;--}_\lambda E^{\mu}_{\alpha}=
E^{\mu}_{\alpha}\;N^{--} \label{5.48.4}\ee
\bea
N^{--} \equiv {1\over 4n}\;[\;\partial^{\;-}_\mu
H^{--+}+\partial^{\;+}_\mu H^{-3\mu}-
\Delta^{--}\ln(\det \partial H) -2n\;\Delta^{--}\ln\psi \;]\;.
\label{5.48.5}\eea

Thus, ${\cal D}^{\;--}_{\lambda}E^{\mu}_{\alpha}$, the last $\lambda$-world
geometric object to be specified, also turned out expressed in terms of the
quaternionic potentials.

To be convinced that eq. \p{5.48.4} is correct, one may compare the
transformation properties of its both sides. It is not difficult to
find
\bea
\delta {\cal D}^{\;--}_\lambda E^{\mu}_{\alpha} &=&
 \lambda_\alpha{}^\beta \;{\cal D}^{\;--}_\lambda E^{\mu}_{\beta}+
({\cal D}^{\;--}_\lambda E^{\rho}_{\alpha})\;
\partial^{\;+}_\rho\lambda^{-\mu}-
 \Delta^{--}\lambda\; E^{\mu}_{\alpha} \nn \\
&& +E^{\mu}_{\alpha}\;\Delta^{--}
[\;{1\over 2n}\;(\partial^{\;+}_\rho\lambda^{-\rho})-
\lambda\; ]+
(\Delta^{--}\lambda^{++})\;{\cal D}^{--}_\lambda E^{\mu}_{\alpha} \;,
\label{5.48.6}\eea
where we have used the property
\bea
{\cal D}^{++}\;[\;(z^0_A)^{-2}\;
\psi\;\Delta^{--}\partial^{\;+}_\nu\lambda^{-\mu}\;] &=&
-(z^0_A)^{-2}\;\psi\;(\Delta^{--}\lambda^{++})\delta_\nu^\mu \nn\\
\Rightarrow \;\;\;\Delta^{--}\partial^{\;+}_\nu\lambda^{-\mu} &=&
{1\over 2n}\;\Delta^{--}\;(\partial^{\;+}_\rho\lambda^{-\rho})\;\delta_\nu^\mu
\label{5.48.7}\eea
(recall that in the HK case \cite{a2} the r.h.s. of the analogous identity is
zero). Directly evaluating the transformation properties of
different pieces in $N^{--}$, one finds
\bea
\delta N^{--}=(\Delta^{--}\lambda^{++})\;N^{--}+
{1\over 2n}\;\Delta^{--}(\partial^{\;+}_\rho\lambda^{-\rho})
-\Delta^{--}\lambda
\label{5.49.1}\eea
that suggests the same transformation law for both sides of \p{5.48.4}.
\subsection{The quaternionic metric in the $\lambda$-basis}

Summarizing the computations of the previous subsections, one may
write the covariant derivatives of the quaternionic geometry (their
parts involving no derivatives with respect to the coordinates $z^{\pm\pm}$)
in the following form
\bea
{\cal D}^{\;+}_\alpha &=& {1\over \psi(1-{\cal L}H)}\;e_\alpha{}^\rho
(\partial{\hat H}^{-1})_\rho{}^\mu\partial^{\;+}_\mu \equiv
E^{+\mu-}_\alpha\partial^{\;+}_\mu
\label{5.49.1'}\\
{\cal D}^{\;-}_\alpha &=& (z^0_A)^{-2}\;\{[\;\psi E_\alpha{}^\mu N^{-2}+
{\cal L}^+_\alpha H^{-3\mu}-\psi E_\alpha{}^\rho\;\partial^{\;+}_\rho H^{-3\mu}
+z^{--}_AE_\alpha{}^\mu\;]\;\partial^{\;+}_\mu \nn \\
&& -e_\alpha{}^\mu\;\hat\partial^{\;-}_\mu -
\rho^{-}\;(z^0_A{\partial\over\partial z^0_A})
-\Delta^+_\alpha\;(\psi\;\omega^{--})\} \nn\\
&\equiv &  E_\alpha{}^{-\mu-}\;
\partial^{\;+}_\mu+E_\alpha{}^{-\mu+}\;\hat\partial^{\;-}_\mu+
\omega^{-}_{\alpha}\;Z^{\;0}-
\omega^-_{\alpha(\sigma\rho)}\;\Gamma^{(\sigma\rho)}\;.
\label{5.49.2}\eea
Defining
\be
g^{MN}_{(\lambda)}=-E^{+M\alpha}E^{-N}_\alpha
+E^{-M\alpha}E^{+N}_\alpha\;, \;\;\;\;M,N=(\nu+,\;\nu-)\;,\label{5.49.3}
\ee
one obtains
\be
g^{\mu+\nu+}_{(\lambda)}=0 \label{5.50.1a}\ee
\bea
g^{\mu+\nu-}_{(\lambda)} &=& g^{\nu-\mu+}_{(\lambda)}=
E^{-\mu+\;\alpha}E^{+\nu-}_\alpha= R\;T^{\mu\rho}\;
(\partial{\hat H}^{-1})_\rho{}^\nu \label{5.50.1b} \\
g^{\mu-\nu-}_{(\lambda)}&=& -E^{+\mu- \;\alpha}E^{-\nu-}_\alpha+
E^{-\mu-\;\alpha}E^{+\nu-}_\alpha\nn \\
&=& -2R\;T^{\rho\lambda}\;(\partial{\hat H}^{-1})_\lambda{}^\sigma
(\partial{\hat H}^{-1})_\rho{}^{(\mu}\;\partial^{\;+}_\sigma{\hat
H}^{-3\nu)}\;.
\label{5.50.1c}\eea

We stress that the dependence on $z^{--}_A, z^0_A$ present in \p{5.49.1},
\p{5.49.2} completely drops out from the $\lambda$-basis metric which is
given on the manifold $\{x^+_A,\;x^-_A,\;w^{\pm i}\}$. Of course, one may
easily check that the metric possesses, as in the HK case, the property of
the covariant independence of harmonics. It is also to the point here
to mention that the bridge from the $Sp(n)-\tau$-frame to the
$\lambda$-frame is expressed in terms of the harmonic connection
$\omega^{++}$ by the same relation as in the HK case, so we do not give
it here.

\subsection{Gauges}

Similarly to the HK case, in practice it is advantageous to fix, in
one or another way, the gauges with respect to different
$\lambda$-transformation, thereby decreasing the number of
independent quantities originally present in the theory. Some
gauges (eqs. \p{5.30.1}, \p{5.30.2}, \p{5.43.1}, \p{5.43.3}) have been
already imposed in the process of solving quaternionic geometry constraints.
Here we enforce further gauge-fixing analogous to the one employed in
the HK case \cite{a2}.

First, one may put $([R]=cm^{-2})$
\be
{\cal L}^+_\mu=R\;\Omega_{\mu\nu}\;x^{+\nu}_A\equiv R\;x^+_{A\mu} \;\;
\Rightarrow \;\;
 T_{\mu\nu}=- R\Omega_{\mu\nu}  \label{5.51.2} \ee \be
\lambda^+_\mu+\hat\partial^{\;-}_\mu\lambda^{+\nu}x^+_{A\nu}+R^{-1}
\hat\partial^{\;-}_\mu\lambda^{++} = 0 \;\Rightarrow \;
\lambda^+_\mu=-{1\over 2}R^{-1}\hat\partial^-_\mu\tilde\lambda^{++},\;
\tilde\lambda^{++}\equiv \lambda^{++}+R\lambda^{+\nu}x^+_{A\nu}.
\label{5.51.3} \ee
 Note the relation
\be
\hat\partial^{\;-}_{[\rho}\lambda^+_{\mu]}=-{1\over 2}\;R^{-1}\;T_{\rho\mu}
\partial^{\;--}\tilde\lambda^{++}={1\over 2}\;\Omega_{\rho\mu}\;
\partial^{\;--}\tilde\lambda^{++}\;. \label{5.51.4} \ee

In this gauge the transformation law of the remaining (non-removable)
prepotential ${\cal L}^{+4}$ is greatly simplified
\be
\delta {\cal L}^{+4}=-\partial^{\;++}\tilde\lambda^{++}-
(\partial^{\;--}\tilde\lambda^{++})\;{\cal L}^{+4} \label{5.51.5} \ee
or, in the active form
\bea
\delta^* {\cal L}^{+4}(x^+_A,w^{\pm i})&\simeq&
{\cal L}^{+4'}(x^+_A,w^{\pm i})-{\cal L}^{+4}(x^+_A,w^{\pm i})
= -\partial^{\;++}\tilde\lambda^{++}-
(\partial^{\;--}\tilde\lambda^{++})\;{\cal L}^{+4} \nn \\
&& + \tilde\lambda^{++}\;\partial^{\;--}{\cal L}^{+4}+
{1\over 2}R^{-1}\;\hat\partial^{\;-\mu}\tilde\lambda^{++}\;
\hat\partial^{\;-}_{\mu}
{\cal L}^{+4}\;. \label{5.51.6} \eea

Note that the isometries of quaternionic metric correspond to
\be
\delta^* {\cal L}^{+4}=0\;. \label{5.51.7}
\ee
Denoting the relevant transformation parameter in \p{5.51.6} by $K^{++}$,
$$
K^{++} \equiv K^{++}_A c_A,
$$
with $c_A$ being {\it constant} infinitesimal parametries of the isometries,
one obtains
from \p{5.51.6} the equation for $K^{++}$
\be
\partial^{\;++} K^{++}+
(\partial^{\;--} K^{++})\;{\cal L}^{+4}
 - K^{++}\;\partial^{\;--}{\cal L}^{+4}-
{1\over 2}R^{-1}\;\hat\partial^{\;-\mu}K^{++}\;\hat\partial^{\;-}_{\mu}
{\cal L}^{+4}=0\;.
\label{5.51.6a} \ee
(and the same for $K^{++}_A$).
The quantity $K^{++}_A$ encodes all the
information about isometries of the quaternionic manifold with the given
potential ${\cal L}^{+4}$ and can naturally be called quaternionic Killing
potential \cite{a11}.

Transformations \p{5.51.5}, \p{5.51.6} and the equation for Killing
potential \p{5.51.6a} have been deduced
earlier in \cite{a11} starting with the harmonic superspace action of $N=2$
matter in $N=2$ SG background. Here we rediscovered the same
objects and relations on the pure geometrical ground, starting from the
unconstrained prepotential formulation of quaternionic geometry.
One-to-one correspondence between this formulation of quaternionic geometry
and the harmonic superspace description of $N=2$ matter coupled to
$N=2$ SG will be discussed in Sect. 10.

One more gauge also having a prototype in the HK case allows one to
``solder'' the $\lambda$-frame $Sp(n)$ transformations with the
diffeomorphisms of the analytic subspace
\bea
{\tilde e}_\alpha{}^\mu=\delta_\alpha^\mu\;\; \Rightarrow
\lambda_{\alpha\beta}+\lambda\;\Omega_{\alpha\beta}+
\hat\partial^{\;-}_\alpha\lambda^+_\beta=0 \label{5.52.1} \eea
\bea
 \Rightarrow \;\; \lambda_{\alpha\beta}=-
\hat\partial^{\;-}_{(\alpha}\lambda^+_{\beta)}
  \;\;\; \;\; \lambda={1\over 2}\;(\partial^{\;--}\tilde\lambda^{++})\;.
\label{5.52.2b} \eea
One may see that \p{5.52.2b} completely agrees with the previously
imposed gauges \p{5.43.1} ${\tilde B}={\tilde e}$ and ${\tilde
B}=$const or ${\tilde e}=$const. These gauges now coincide in view of
\p{5.52.1}.

\setcounter{equation}{0}
\section{Example: homogeneous manifold
$Sp(n+1)/Sp(1)\times Sp(n)$}

Here we will consider the simplest example of $4n$ dimensional quaternionic
manifold corresponding to the choice ${\cal L}^{+4}=0$. As opposed to the HK
case where such an option results in a trivial flat manifold, in the case
under consideration we are left with a curved homogeneous space
$Sp(n+1)/Sp(1)\times Sp(n)$
(or  $Sp(n,1)/Sp(1)\times Sp(n)$, for the negative value of the parameter $R$)
which can
thus be regarded as the ``maximally flat'' connected $4n$ dimensional
quaternionic manifold.

The choice
\be
{\cal L}^{+4}=0 \label{5.54.1} \ee
in the Killing potential equation \p{5.51.6a}
reduces the latter to the following one (we assume that all the gauges
employed above are imposed)
\be
\partial^{++}K^{++}(x^+,w)=0\;,
\label{5.54.2} \ee
the general solution of which is
\bea
K^{++} &=& w^{+i}w^{+j}\;c_{(ij)}+x^{\mu+}x^{\nu+}\;c_{(\mu\nu)} +
x^{\mu+}w^{+i}\;c_{\mu i}\label{5.54.3} \\
\lambda^+_\mu&=&-
{1\over 2}\;R^{-1}\;(\partial^{\;-}_\mu+
R\;x^+_\mu\;\partial^{\;--})K^{++}) \nn\\
&=&-{1\over 2}\;R^{-1}\;[\;w^{+i}c_{\mu i}+2\;x^{+\nu}c_{(\mu\nu)}+
R\;x^+_\mu\;(2w^{+i}w^{-j}c_{(ij)}+x^{\mu+}w^{-i}c_{\mu i})\;]\;,
\label{5.55.1} \eea
whence
\be
\lambda^{++}=K^{++}-R\;\lambda^{+\nu}x^+_\nu+
w^{+i}w^{+j}\;c_{(ij)}+{1\over 2}\;x^{\mu+}w^{+i}c_{\mu i}\;.
\label{5.55.2} \ee

Studying the Lie algebra of the variations
\be
\delta x^{\mu+}=\lambda^{+\mu},\;\;\delta w^{-i}=0,\;\;
\delta w^{+i}=\lambda^{++}w^{-i}\;, \label{5.55.3} \ee
one finds that it coincides with that of $Sp(n+1)$ (or with $Sp(n,1)$ for
$R<0$ in \p{5.55.1}), the parameters $c_{(ij)},\; c_{(\mu\nu)},\;c_{\mu i}$
being associated, respectively, with the generators of the subgroups $Sp(1)$,
$Sp(n)$ and with those of the coset $Sp(n+1)/ Sp(1)\times Sp(n)$. It is easy
to show that in this case
\be
H^{+3\nu}=0\;,\;\;H^{+4}=0\;,\;\; H^{--+\mu}=x^{-\mu}\;,\;\;
H^{-3\mu}=0        \label{5.55.4c} \ee
\bea \Delta^{\pm\pm} &=&
\partial^{\;\pm\pm}+x^{\pm\rho}\partial^{\;\pm}_\rho\;,\;\;
   1-{\cal L}H=1+R\;x^{+\mu}x^{-}_\mu \equiv 1+ {1\over 2}\;R\;x^2 \nn \\
\rho^-_\mu &=& {1\over 1+ {1\over 2}\;R\;x^2}\;x^{-}_\mu\;,\;\;
E_\alpha{}^\mu=(1+ {1\over 2}\;R\;x^2)\;
(\delta_\alpha^\mu-R\;x^+_\alpha x^{-\mu})
 \label{5.55.4e}\eea \be
E_{\mu\beta} = {1\over 1+ {1\over 2}\;R\;x^2}\;\left( \Omega_{\beta\mu}
+R\;{x^+_\mu x^-_\beta\over 1+ {1\over 2}\;R\;x^2} \right)
 \label{5.55.4f}\ee
\be
\omega^{++}_{\alpha\beta}=R\;x^+_\alpha x^+_\beta\;, \;\;\;
\omega^{--}_{\alpha\beta}={R\over 1+ {1\over 2}\;R\;x^2}\;
x^-_\alpha x^-_\beta\;. \label{5.55.4h}\ee

It is also easy to check the covariant independence of
$E_\alpha{}^\mu$ of the harmonics
\bea
{\cal D}^{\;--}E_\alpha{}^\mu &=&
(x^{-\rho}\partial^{\;-}_\rho)\;E_\alpha{}^\mu
+(\omega^{--})_\alpha{}^\beta \;E_\beta{}^\mu \nn \\
&=& (1+ {1\over 2}\;R\;x^2)\;x^-_\alpha x^{-\mu}-x^-_\alpha  x^{-\mu}-
{1\over 2}\;R\;x^-_\alpha  x^{-\mu}\;(x^2)=0\;, \label{5.56.1}\eea
as well as to verify all the remaining general relations
derived earlier, to find the induced
group parameters $\lambda,\; \lambda^{-\mu}$, etc.
The nonvanishing component of the $\lambda$-world metric is given by the
following simple expression
\be
g^{\mu+\nu-}= - (1+{R\over2}\;x^2)
(\Omega^{\mu\nu}-R\;x^{+\mu}x^{-\nu}) \;. \label{5.56.2}\ee

The bridges can be shown to vanish in the present case, so $w^{\pm i}_A=
w^{\pm i}$ and $\lambda^{++}=\tau^{++}$, and also
\be
x^{\mu i}= -x^{\mu +}w^{-i}+x^{\mu-}w^{+i}\;.
\label{5.56.3}\ee
To find the $\tau$-world metric, one should convert the
indices $\mu,\;\nu$ in \p{5.56.2} with the matrices
\bea
\hat\partial^{\;-}_\mu x^{\rho k}=
-\delta_\mu^\rho \;w^{-k}+R\;x^+_\mu x^{-\rho}w^{-k},
\;\; \partial^{\;+}_\nu  x^{\sigma i}=\delta_\nu^\sigma \;w^{+i}
\label{5.56.4}\eea
and symmetrize indices $\rho k,\; \sigma i$ (without factor 1/2). One gets
\bea
g^{\rho k, \sigma i}&=&
\epsilon^{ki} (1+{R\over2}\;x^2)
(\Omega^{\rho\sigma}-R\;x^{\rho j}x^\sigma_j)
\;, \nn\\
g_{\sigma i,\rho k}&=&{\epsilon_{ik}\over 1+{R\over2}\;x^2}
\left( \Omega_{\sigma\rho}-
R\;{x^j_\sigma x_{\rho j}\over 1+{R\over 2}\;x^2} \right)\label{5.57.1}\eea
that coincides with the metric of the space  $Sp(n+1)/Sp(1)\times Sp(n)$
in the appropriate projective coordinates (we work with the standard
dimension $1$ coordinates).

Finally, we leave it to the reader to reproduce all the formulas given here
starting with the standard nonlinear realization description of the coset
space $Sp(n+1)/Sp(1)\times Sp(n)$, defining the corresponding Cartan forms,
etc. In order to do this unambigously, one needs to extend this space to
the harmonic one
by introducing the harmonics $u^{\pm i}$ on the  extra automorphism $SU(2)$
(acting on the $Sp(1)$ indices of the $Sp(n+1)$ generators) and then
to identify the analytic subspace $\{x^+_\mu, w^{\pm i}\}$ with the
following coset of $Sp(n+1)\times SU(2)_A$
\bea
\{x^+_\mu, w^{\pm i}\}={Sp(n+1)\times SU(2)_A \over
(L^{(\mu\nu)},P^{\mu+},I^{++},I^0, I^{--}-T^{--},T^0)},\label{5.58.1}\eea
where $I$ and $T$ denote the generators of the subgroups
$Sp(1)\subset Sp(n+1)$ and  $SU(2)_A$, respectively, $L^{(\mu\nu)}$ stands
for the generators of $Sp(n)\subset Sp(n+1)$ and $P^{\mu+}$ for those of
the coset $Sp(n+1)/Sp(1)\times Sp(n)$ generators $P^{\mu i}$ which commute
with  $I^{++}, I^{--}-T^{--}$ (their number equals $2n$).

In Sect. 11 we will discuss less trivial examples of homogeneous
quaternionic manifolds with ${\cal L}^{+4} \neq 0$.

\setcounter{equation}{0}
\section{HK manifolds as a contraction of the quaternionic ones}

Here we show that in the limit $R\rightarrow 0$ the formulation of quaternionic
geometry given in the previous Sections goes over to that of the HK one with
the central charge operators included \cite{a2}.

Let us begin with the $\tau$-world constraints and consider first what
happens
in this limit with the bi-harmonic space $\{x^{\mu i},\; w^{\pm i},\;
z^{++},\;z^{--},\; z^0\}$. Rescaling $z$-coordinates as
\be
z^{\pm\pm}\equiv R \;{\tilde z}^{\pm\pm}, \;\;z^0\equiv exp\{ R{\tilde z}^0 \}
\label{5.65.1}\ee
and defining
\be
{\tilde Z}^{\;\pm\pm}\equiv R\;
Z^{\;\pm\pm},\;\;{\tilde Z}^{\;0} \equiv R\;{\tilde Z}^{\;0}
\label{5.65.2}\ee
one finds that in the limit $R\rightarrow 0$ the expressions \p{5.10.1},
\p{5.10.2} become
\be
{\tilde Z}^{\;0} = {\partial\over\partial{\tilde z}^0}\;,\;\;\;
{\tilde Z}^{\pm\pm}={\partial\over\partial {\tilde z}^{\mp\mp}}\;,\;\;\;
w^{\pm i}=u^{\pm i} \ee
\be
[\;{\tilde Z}^{\;0},\;{\tilde Z}^{\;\pm\pm}  \;]
= 0\;,\;\;\;\;[\; {\tilde Z}^{\;++},\;{\tilde Z}^{\;--}\;] = 0
\label{5.66.2} \ee
\bea
{\cal D}^{\;++}&=&\partial^{\;++}_u+{\tilde z}^{++}
{\partial\over\partial{\tilde z}^0}
+2\;{\tilde z}^0{\partial\over\partial{\tilde z}^{--}}\nn \\
{\cal D}^{\;--}&=&\partial^{\;--}_u+2\;{\tilde z}^{0}
{\partial\over\partial{\tilde
z}^{++}} +{\tilde z}^{--}{\partial\over\partial{\tilde z}^{0}}\nn \\
{\cal D}^{\;0} &=&\partial^{\;0}_u+2\;
({\tilde z}^{++}{\partial\over\partial{\tilde
z}^{++}} -{\tilde z}^{--}{\partial\over\partial{\tilde z}^{--}})
 \;. \label{5.66.1}
\eea

Thus, ${\tilde Z}^{++},\;{\tilde Z}^{--},\;{\tilde Z}^0$
become just the central
charge operators introduced in \cite{a2} as a useful device for solving the
HK geometry constraints in a regular way. The
bi-harmonic space goes over to the standard twistor-harmonic extension of the
$\tau$-basis HK manifold enlarged by the extra central charge coordinates
\be
\{x^{\mu i}, w^{\pm j}, z\} \;\;
\Rightarrow \{x^{\mu i}, u^{\pm j}, {\tilde z}\}\;.
\label{5.66.3}\ee

Further, the quaternionic geometry defining constraint
\p{5.19.2} formally preserves its form in the
limit $R\rightarrow 0$
\bea
[\;{\cal D}^{\;+}_\alpha,\;{\cal D}^{\;+}_\beta\;]_\rho{}^{\rho'}
=- 2\;\delta_\rho^{\rho'}
\Omega_{\alpha\beta}{\tilde Z}^{\;++}\;,\;\;\;
{\cal D}^{\;+}_\alpha=u^{+i}
{\cal D}_{\alpha i}\;, \;\;\;
{\tilde Z}^{\;++}=-u^{+i}u^{+j}Z_{(ij)}\;. \label{5.66.4}
\eea
However, because the holonomy $Sp(1)$ contracts into the trivial algebra of
three flat ${\tilde z}$-translations \p{5.66.2}, the status of \p{5.66.4}
radically changes. Namely, ${\cal D}^{\;+}_\alpha$
has now the following structure
(cf. \p{5.22.3})
\be
{\cal D}^{\;+}_\alpha={\hat E}_\alpha^{+\mu k}\partial_{\mu k}-
\hat\omega^+_{\alpha(\rho\sigma)}\;\Gamma^{\rho\sigma}
-\hat\omega_\alpha^{+-+}\;{\partial\over\partial{\tilde z}^0}-
\hat\omega_\alpha^{+--}\;{\partial\over\partial{\tilde z}^{--}}-
\hat\omega_\alpha^{+3}\;{\partial\over\partial{\tilde z}^{++}}\;,
\label{5.67.1}\ee
where hat means that we consider the $R=0$ contraction of the corresponding
objects of the quaternionic geometry. Note that \p{5.67.1} contains no
${\tilde z}^0$ dependence because $z^0=exp\{ R{\tilde z}^0 \} \rightarrow 1$ as
$R\rightarrow 0$. It is easy to see that \p{5.66.4} produces for
${\hat E}_\alpha^{+\mu k}$, $\hat\omega^+_{\alpha(\rho\sigma)}$ the standard
HK constraints while in the quaternionic case the corresponding constraints
essentially involved the vielbeins multiplying the $z$-derivatives. Thus,
in the $\tau$-world eq. \p{5.66.4} (accompanied by the the evident harmonic
commutation relations) amounts to the familiar constraints of HK geometry.

When passing to the $\lambda$-world, we have to rescale $z^0_A,\;z^\pm_A$ in
the
same fashion as in eq. \p{5.65.1} and simultaneously to rescale the bridges
and the gauge group parameters. There is a minor difference of the resulting
 $\lambda$-world HK constraints as compared with the corresponding ones
obtained in \cite{a2}. The difference stems from the fact that in the
quaternionic
case we are led to include $z^0_A$ into the analytic subspace (together with
$z^{++}_A$) in order to have analytic $Sp(1)$ weight transformations. No any
$Sp(1)$ weight structures survive in the $R=0$ contraction limit, so one can,
in principle,
restrict oneself to considering the analytic space involving only
${\tilde z}^{++}_A$ (treating ${\tilde z}^{0}_A$ on equal footing with
${\tilde z}^{--}_A$) and this is just what has been done in \cite{a2}.
Contraction of the
quaternionic geometry relations yields another version of the central charge
modified
$\lambda$-world HK constraints, with ${\tilde z}^{0}_A$ included into the set
of analytic space coordinates. The final results, however, are the same as in
the version worked out in \cite{a2}. Without entering into details, we only
mention
the basic redefinitions one should make to ensure an unambigous passing to the
limit $R\rightarrow 0$ in the $\lambda$-world relations deduced in the previous
Section
\bea
H^{+4}=R\;{\tilde H}^{+4},\;\; {\cal L}^+_\mu=R\;{\tilde {\cal L}}^+_\mu,\;\;
{\cal L}^{+4}=R\;{\tilde {\cal L}}^{+4}\;. \label{5.68.1} \eea

Finally, let us quote some basic objects of quaternionic geometry in the
HK limit $R\rightarrow 0$
\bea
H^{+3\nu}&\Rightarrow &\;-{1\over 2}\;{\tilde T}^{\nu\mu}
\;(\partial^{\;-}_\mu {\tilde {\cal L}}^{+4}-\partial^{\;++}{\tilde {\cal
L}}^+_\mu) \nn \\
\omega^{++}_{\alpha\beta} &\Rightarrow &
 \; e_{(\alpha}{}^\mu\;\Delta^{++}e_{\mu\beta)}
+e_{(\alpha}{}^\nu\;\partial^{\;-}_\nu H^{+3\mu}\;e_{\mu\beta)} \nn \\
g^{\mu+\nu-}_{(\lambda)} &\Rightarrow &\;-{\tilde T}^{\mu\rho}
(\partial H^{-1})_\rho{}^\nu \label{5.69.1} \\
g^{\mu-\nu-}_{(\lambda)} &\Rightarrow & \;2\;{\tilde T}^{\rho\lambda}
(\partial H^{-1})_\lambda{}^\sigma (\partial H^{-1})_\rho{}^{(\mu}
\;\partial^{\;+}_\sigma H^{-3\nu )}\;, \;\;\;\;\;{\rm etc.}\nn \eea
Thus, as it should be, in the limit when the $Sp(1)$ curvature vanishes, one
self-consistently recovers the unconstrained harmonic space formulation of HK
manifolds.

\setcounter{equation}{0}
\section{Geometry of $N=2$ matter in $N=2$ supergravity background}

Here we demonstrate that the equations of motion following from the general
harmonic superspace off-shell action of
$N=2$ matter coupled to $N=2$ supergravity
given in \cite{a11} have a simple interpretation in terms of quaternionic
geometry in the
unconstrained harmonic space formulation presented above. Surprisingly, the
components
\be
\omega=u^-_iq^{+i},\;\;N^{++}={u^+_iq^{+i}\over u^-_jq^{+j}}
\label{5.59.1}\ee
of the supergravity hypermultiplet
compensator $q^{+i}$ acquire a clear geometric meaning
as the coordinates $z^0_A, z^{++}_A$ of the analytic space
$\{z^0_A, z^{++}_A,w^{\pm i},x^{\mu+}\}$ of the quaternionic geometry, while
$x^{\mu+}$ (like in the HK case) are identified with the $N=2$ matter
hypermultiplet superfileds.

Let us start with the $N=2$ matter - supergravity  action \cite{a10} written
in terms of compensators $\omega,\;N^{++}$ \p{5.59.1} \cite{a15}
\bea
S^{\scriptstyle N=2}_{\scriptstyle SG+matter} &=&
-{1\over 2\kappa^2}\int d\zeta^{(-4)}du \;\omega^2\{  {\cal H}^{+4}-
{\cal D}^{\;++}_u N^{++}-(N^{++})^2 \nn \\
&& -{\kappa^2\over\xi^2}[{\tilde{\cal
L}}^{+4}(Q,w)+{\tilde{\cal L}}^{+\mu}(Q,w){\cal D}^{\;++}_wQ^{+}_{\mu}]\}
\label{5.60.1}\eea \bea w^{-i}=u^{-i}\;,\;\; w^{+i}=u^{+i}-N^{++}u^{-i}\;,\;\;
{\cal D}^{\;++}_w={\cal D}^{\;++}_u+N^{++}{\cal D}^{\;0}\;.
\label{5.60.2}\eea
In \p{5.60.1}, integration goes over the analytic subspace of harmonic
$N=2$ superspace, $Q^{+}$ are $N=2$ matter hypermultiplet analytic superfields,
${\cal H}^{+4}$ is the analytic harmonic vielbein of conformal $N=2$
supergravity, ${\cal D}^{\;++}_{u}$ is the supergravity-covariantized
harmonic derivative with respect to $u^{\pm i}$ in the analytic basis of
$N=2$ harmonic superspace and $\kappa$, $\xi$ are, respectively, Newton
and sigma-model coupling constants (for details see ref.
\cite{{a3},{a10},{a11},{a15}})\footnote{The full matter-supergravity action
also includes the purely supergravity part, which, however, is irrelevant
here to our purposes.}.

The action \p{5.60.1}
is invariant \cite{a11} under two kinds of gauge transformations with analytic
parameters: diffeomorphisms (the parameters $\lambda^{+\mu}(Q,w)$) and the
quaternionic analog of the HK and K\"ahler transformations (the parameters
$\Lambda^{++}(Q,w)$):
\bea
\delta Q^{+\mu}=\lambda^{+\mu}(Q,w)\label{5.60.3}\eea
\bea
\delta N^{++} &=& -{\kappa^2\over\xi^2}\;\Lambda^{++}\;,\;\;\;
\delta\omega={1\over
2}\;{\kappa^2\over\xi^2}\;[-\partial^{\;--}_{w}\Lambda^{++}
-{\tilde{\cal L}}^+_\mu\partial^{\;--}_w\lambda^{+\mu}]\omega
 \label{5.60.4}\\\delta{\tilde{\cal L}}^+_\mu
 &=& -\hat\partial^{\;-}_\mu\Lambda^{++}-
{\tilde{\cal L}}^+_\rho\;\hat\partial^{\;-}_\mu\lambda^{+\rho}\;,\;\;\;
\delta{\cal D}^{\;++}_w=-{\kappa^2\over\xi^2}\;
\Lambda^{++}\;{\cal D}^{\;0} \nn \\
\delta{\tilde{\cal L}}^{+4} &=& \partial^{\;++}_w\Lambda^{++}
+{\tilde{\cal L}}^+_\mu\partial^{\;++}_w\lambda^{+\mu} \nn \\
&& + {\kappa^2\over\xi^2}\;[\;
\partial^{\;--}_w\Lambda^{++}\;{\tilde{\cal L}}^{+4}
+\partial_w^{--}\lambda^{+\mu}\;{\tilde{\cal L}}^{+}_{\mu}
{\tilde{\cal L}}^{+4} -
\Lambda^{++}\;{\tilde{\cal L}}^+_\mu
Q^{+\mu}\;]
\label{5.60.5}\eea
\bea
\hat\partial^{\;-}_\mu\equiv \partial^{\;-}_\mu-{\kappa^2\over\xi^
2}\;
{\tilde{\cal L}}^+_\mu\;\partial^{\;--}_w \;.
\label{5.60.6}\eea

Identifying
\bea
Q^{+\mu}&\equiv & x^{+\mu}_A\;,\;\;\;
{\kappa^2\over\xi^2}\;\Lambda^{++}\equiv -\lambda^{++}\;,\;\;\;
{\kappa^2\over\xi^2}\;{\tilde{\cal L}}^+_\mu\equiv -{\cal L}^+_\mu\; \nn \\
{\kappa^2\over\xi^2}\;{\tilde{\cal L}}^{+4}
 &\equiv & {\cal L}^{+4}\;,\;\; \omega \equiv z^0_A\;,\;\;
N^{++} \equiv z^{++}_A\;,\;\;
{\kappa^2\over\xi^2} \equiv - R\;, \label{5.61.1}\eea
we observe the one-to-one correspondence between \p{5.60.3} - \p{5.60.6} and
the quaternionic geometry transformation laws in the gauge
${\tilde e}={\tilde B}$ \p{5.43.3} (eqs. \p{5.26.2}, \p{5.38.1}, \p{5.40.3}).
We see that this correspondence requires $R< 0$ i.e. in the case at hand,
while considering homogeneous quaternionic manifolds, one necessarily deals
with their {\it noncompact} versions (see Sect. 11).
Recalling \p{5.4.4}, one
recovers the familiar relation \cite{{a9},{a14}} between the coupling constant
${\kappa^2\over\xi^2}$ and the scalar curvature of the quaternionic manifold of
$N=2$ matter
\be
R_{\alpha i\;\beta j}^{\phantom{\alpha i\;\beta j}\alpha i\;\beta j}
=- 8n(n+2){\kappa^2\over\xi^2}\;.
\label{5.61.2}\ee

Thus the basic entities of the most general $N=2$ matter action in
the supergravity background, ${\tilde{\cal L}}^{+}_{\mu}$ and
${\tilde{\cal L}}^{+4}$, are recognized as the unconstrained quaternionic
geometry potentials, the matter superfields $Q^{+}$ being the coordinates
of the analytic subspace of harmonic extension of the target quaternionic
manifold. This makes evident the one-to-one correspondence between off-shell
locally $N=2$ supersymmetric sigma models and quaternionic manifolds revealed
at the component on-shell level in \cite{a9}.

Let us now, as we promised, give the geometric interpretation to the
equations of motion following from \p{5.60.1}.

Varying \p{5.60.1} with respect to
$\omega$, $N^{++}$ and $Q^{+\mu}$ yields, respectively,
\bea
\delta\omega: \;\;\;&&{\cal D}^{\;++}_u
N^{++}+(N^{++})^2- {\cal H}^{+4}+({\cal L}^{+4}
+{\cal L}^+_\mu\;{\cal D}^{\;++}_wQ^{+\mu})= 0
\label{5.62.1}\\
\delta N^{++}:\;\;\;&& N^{++}-{\cal D}^{\;++}_u\ln\omega-{1\over 2}\Gamma^{++}
+{1\over 2}\;{\cal L}^+_\mu\; Q^{+\mu} \nn \\
&& - {1\over 2}(\partial^{\;--}_w{\cal L}^{+4}+\partial^{--}_w{\cal L}^+_\mu\;
{\cal D}^{\;++}_wQ^{+\mu}) = 0
\label{5.62.2} \\
\delta Q^{+\mu}:\;\;\;&& \partial^{\;-}_{\mu} {\cal L}^{+4}+
(\partial^{\;-}_{\mu}{\cal L}^{+}_{\rho} -
\partial^{\;-}_{\rho}{\cal L}^{+}_{\mu})\;{\cal D}^{\;++}_{w} Q^{+\rho}
-\partial^{\;++}_w{\cal L}^+_\mu  \nn \\
&& +2\; {\cal L}^+_\mu \;(N^{++}-
{\cal D}^{\;++}_u\ln\omega-{1\over 2}\Gamma^{++}) \nn\\
&& -[\; {\cal H}^{+4}-{\cal D}^{++}_u N^{++}-(N^{++})^2\;]\;\partial^{\;--}_w
{\cal L}^+_\mu=0\;.
\label{5.62.3}\eea
Here $\Gamma^{++}$ is the $L(1)$ connection of conformal $N=2$ supergravity
(it is expressed in a proper way through the analytic $N=2$ supergravity
prepotentials, its precise form is of no interest for us)
and we passed to renormalized objects \p{5.61.1}. Substituting \p{5.62.1},
\p{5.62.2} into eq. \p{5.62.3}, one recasts the
latter into the form
\bea
(\hat\partial^{\;-}_\mu{\cal L}^+_\rho-\hat\partial^{\;-}_\rho{\cal
L}^+_\mu){\cal D}^{\;++}_wQ^{+\rho}=-\hat\partial^{\;-}_\mu{\cal L}^{+4}
+{\cal L}^+_\mu({\cal L}^+_\rho Q^{+\rho})+\partial^{\;++}_w{\cal L}^+_\mu
+{\cal L}^{+4}\partial^{\;--}_w{\cal L}^+_\mu
\label{5.62.4}\eea
that is precisely the relation \p{5.41.1} upon the identification
\be
{\cal D}^{\;++}_wQ^{+\rho}=
{\cal D}^{\;++}_u Q^{+\rho}+N^{++}Q^{+\rho}=H^{+3\rho}\;.
\label{5.62.5}\ee
Recall that the analogous interpretation, but in the framework of the harmonic
space formulation of HK geometry, can be given to the $Q^{+}$ equations
following from the general off-shell action of sigma model with rigid $N=2$
supersymmetry \cite{a2}.

It remains to clarify the geometric meaning of the relations \p{5.62.1},
\p{5.62.2} and \p{5.62.5} which involve the objects $N^{++},\;\omega$ having
no prototypes in the harmonic superspace
action of rigid $N=2$ matter. While the identification
\p{5.62.5} has a direct analog in the rigid case (with $N^{++}$ put
equal zero), the remaining two relations are essentially new. Remarkably,
for all of them one can find a natural place within the quaternionic geometry
formulation presented in this paper.

We note first that the counterpart of the
derivative $ D^{\;++}$ \p{5.27.3} is just ${\cal D}^{\;++}_u$: it is invariant
under the $\lambda^{+\mu}$ and $\lambda^{++}$ invariances and transforms
only under gauge group of conformal $N=2$ SG. In other words, it can be
regarded as the  $N=2$ SG gauge covariantization of  $ D^{++}$ \p{5.27.3}.
Then ${\cal D}^{\;++}_w$ corresponds to  ${\cal D}^{\;++}_\lambda$ defined
in \p{5.27.4}.
Let us now consider the action of  $ D^{\;++}$ \p{5.27.3} on $z^{++}_A$:
\be
D^{\;++}z^{++}_A=-[\;H^{+4}+(z^{++}_A)^2\;] \;.\label{5.63.1}\ee
Comparing it with \p{5.62.1} and taking into account the relation \p{5.40.3a}
\be
H^{+4}={\cal L}^{+4}+{\cal L}^+_\mu\;{\cal D}^{\;++}_wQ^{+\mu}=
{\cal L}^{+4}+{\cal L}^+_\mu H^{+3\mu} \label{5.63.2}\ee
and the identifications \p{5.61.1}, one concludes that \p{5.62.1} is nothing
else as the $N=2$ SG covariantization of the quaternionic geometry relation
\p{5.63.1}. One simply adds to $H^{+4}$ the $N=2$ SG vielbein ${\cal H}^{+4}$,
thus ensuring covariance both under the quaternionic and  $N=2$ SG
gauge transformations.

Analogously, eq.\p{5.62.2} can be rewritten as
\bea
{\cal D}^{\;++}_u\omega &=&
-{1\over 2}\;[\;\partial^{\;--}_w{\cal L}^{+4}
+\partial^{\;--}_w({\cal L}^+_\mu{\cal D}^{\;++}_wQ^{+\mu})
-{\cal L}^+_\mu\partial^{\;--}_w({\cal D}^{++}_wQ^{+\mu})-
{\cal L}^+_\mu Q^{+\mu}\;]\omega \nn \\
&& +(N^{++}-{1\over 2}\;\Gamma^{++})\omega \;.
\label{5.64.1}\eea
Comparing it with eq.\p{5.45.3}, we can rewrite it as
\bea
{\cal D}^{\;++}_u\omega=(N^{++}+\phi^{++}-{1\over 2}\;\Gamma^{++})\;\omega
\label{5.64.2}\eea
that is a direct $N=2$ SG covariantization of the relation
\be
 D^{\;++}z^0_A=(z^{++}_A+\phi^{++})\;z^0_A \label{5.64.3}\ee
following from the explicit form of $D^{\;++}$ \p{5.27.3}.

Finally, the relation \p{5.62.5} can also be geometrically interpreted as
the result of action of $D^{\;++}$ \p{5.27.3} on $x^{+\mu}$:
\be
D^{\;++}x^{+\mu}=H^{+3\mu}-z^{++}_A x^{+\mu}\;. \label{5.64.4}\ee

Thus we have proved complete one-to-one correspondence between the
unconstrained
formulation of quaternionic geometry given in the previous Sections and the
equations of motion of general $N=2$ sigma model coupled to $N=2$ SG.

We stress that the $Sp(1)$ coordinates $z^{++}_A=N^{++}$ and $z^0_A=\omega$
play the important role both in the harmonic superspace geometry of
$N=2$ SG and in the harmonic space geometry of $N=2$ matter superfields.
They are common for both geometries and so establish a link between them.

\setcounter{equation}{0}
\section{Potentials for
symmetric quaternionic spaces}

In this Section we quote the explicit form of the potentials ${\cal L}^{+4}$
for all types of connected symmetric quaternionic spaces classified
in \cite{a13} \footnote{A more general class of homogeneous
nonsymmetric quaternionic spaces has been given in \cite{a16}.}.

There are precisely one compact and one noncompact quaternionic cosets for
each simple complex Lie group. Here we specialize to the noncompact
versions of them which one encounters in $N=2$ supergravity.

These cosets are \cite{{a13},{a16}}
\be\begin{array}{cccc}
\fracds{SU(n,2)}{U(n) \times Sp(1)}&\
\fracds{SO(n,4)}{SO(n)\times SU(2) \times Sp(1)}&  \
\fracds{Sp(n,1)}{Sp(n)\times Sp(1)} &
\fracds{G_{2(+2)}}{SU(2)\times Sp(1)}\\[3ex]
\fracds{F_{4(+4)}}{Sp(3)\times Sp(1)}&\
\fracds{E_{6(+2)}}{SU(6)\times Sp(1)}&\
\fracds{E_{7(-5)}}{SO(12)\times Sp(1)}&
\fracds{E_{8(-24)}}{E_7 \times Sp(1)}.
\end{array}
\label{one}
\ee
They have the form $G/H \times Sp(1)$ with $H \in Sp(n)$. The number in
brackets
for the exceptional groups
refers to the chosen real form of the group. It equals to the
difference
between the numbers of noncompact and compact generators for
this real form. The compact case corresponds to the compact real form of the
numerator.

As explained above the coset
$Sp(n,1) \ / Sp(n)\times Sp(1)$ is a ``flat model'' for the quaternionic
geometry. The corresponding $\lpr$ vanishes.
The potentials for the cosets
\be\begin{array}{ccc}
\fracds{SU(n,2)}{U(n) \times Sp(1)}&\
\fracds{SO(n,4)}{SO(n)\times SU(2) \times Sp(1)}&  \
\fracds{G_{2(+2)}}{SU(2)\times Sp(1)}
\end{array}
\label{two}
\ee
were given in \cite{a11}. Here we recall these potentials and
complete the list.
We give only the final expressions for $\lpr$.
The general procedure of finding them based on the analogy with the
Hamiltonian mechanics \cite{Ham} has been recently worked out in \cite{mal}.

The coset parameters $Q^{i\Sigma}$ for manifolds \p{one} have the
quantum numbers of the group generators belonging to the coset: they
always carry the index $i$ ($i = 1,2$) of the fundamental
two-dimensional representation
of the group $Sp(1)$ from the denominator $H\times Sp(1)$ and the index
$\Sigma$ ($\Sigma = 1, 2, ... 2n$) which
refers to the group $H$. The potential $\lpr$ depends
only on the '+'-projection $Q^{+\Sigma}$ of $Q^{i\Sigma}$.
In the sequel we will refer just to $Q^{+\Sigma}$ as the coset parameters.
Clearly, as $H$ defines a linear symmetry on $Q^{+}_{\Sigma}$ and $\lpr$ is
$H$ invariant, one needs to know the assignment of the adjoint
representation of the group in the nominator,
 $G$, with respect to the subgroup $H$.

We briefly sketch how $\lpr$ can be found within the
geometric set-up employed in
previous Sections.
Computations are based on several simplifying observations.

Firstly, since in the above cosets linearly realized $Sp(1)$ acts, the
relevant potentials $\lpr$ do not depend explicitly on the harmonics
$w^{\pm}_{i}$. Then the condition
$${\cal D}^{\;0} \lpr = 4 \;\lpr$$
tells
us that these potentials must be quartic in $Q^{+}_{\Sigma}$ \cite{a11}.

Secondly, the group $G$ defines isometries of these
cosets, so
there exist Killing potentials $K^{++}_A$, the number of which is equal
to $dim \;G$ and which are solutions of eq.
\p{5.51.6a}. For the case at hand this
equation simplifies to
\bea
\partial^{\;++}K^{++}+
(\partial^{\;--}K^{++})\;{\cal L}^{+4} +
{1\over 2}\;\hat\partial^{\;-\mu}K^{++}\;\hat\partial^{\;-}_{\mu}
{\cal L}^{+4}=0 \label{conserv}
\eea
(for simplicity, we put $R=-1$ in accordance with the fact that we consider
here only noncompact manifolds).
The $Sp(1)$ Killing potential $K^{++}_{i\;j}$
has the universal form for all spaces \p{one}
$$  K^{++}_{i\;j} = w^+_i w^+_j  + w^-_iw^-_j \lpr \;,$$
while inspection of eq. \p{conserv} for the symmetries belonging
to the coset and the subgroup $H$ leads to the important equation
\be
Q^+_{\ \Sigma}\; \lpr - \fracds{1}{4} \;\Omega^{\Lambda\Gamma}
\fracds{\partial^{2}  \lpr}{\partial Q^{+\Sigma} \partial Q^{+\Lambda}}\;
\fracds{\partial \lpr}{\partial Q^{+\Gamma}}
 = 0\;.
\label{leq}
\ee
Here $\Omega^{\Sigma\Gamma}$ is the invariant antisymmetric $Sp(n)$ tensor.

It turns out that the equation \p{leq}, together with the requirement of $H$-
invariance of $\lpr$ and the property that $\lpr$ is a quartic polynomial in
$Q^{+}_{\Sigma}$, determine $\lpr$ uniquely: for each case
a $H$-invariant polynomial of degree 4 is unique up to
a numerical normalization factor, which is fixed by
eq. \p{leq}.

\vspace{2 mm}
\noindent
1. $SU(n,2)\ /U(n) \times Sp(1)$.
The coset parameters are
$(Q^+_{\ a},\; \bar{Q}^{+a})$, where $a$ is the index of the fundamental
representation of $U(n)$. The pseudoreal structure is given
by
$$
*Q^+_{\ a} = \bar{Q}^{+a} \;, \;\;\;\; *\bar{Q}^{+a} = - Q^+_{\ a} \;.
$$
The potential $\lpr$ is
\be
\lpr = - (Q^+ \bar{Q}^+)^2\;.
\ee

\vspace{2 mm}
\noindent
2. $SO(n,4) \ /SO(n)\times SU(2)\times Sp(1)$.
The coset parameters
are $Q^+_{\ a\alpha}$ where $a$ is the vector index of $SO(n)$
and $\alpha$ is the spinor index of $SU(2)$.
The pseudoreal structure is given by:
$$
*Q^+_{\ a\alpha} = Q^{+\ \alpha}_{\ a}\;.
$$
The potential is
\be
\lpr = \fracds{1}{2}\; (Q^+_{\ a\alpha}Q^{+\ \alpha}_{\ b} )\;
(Q^{+a}_{\ \ \gamma}Q^{+b\gamma})\;.
\ee

\vspace{2 mm}
\noindent
3. $G_{2(+2)}\ / SU(2)\times Sp(1)$.
Now the coset parameters are
$Q^+_{\ \alpha\beta\gamma}$, where greek indices refer to the spinor
representation of $SU(2)$. They are totally antisymmetric in $\alpha$,
$\beta$, and $\gamma$.
Pseudoreality:
$$
*Q^+_{\ \alpha\beta\gamma} = Q^{+\alpha\beta\gamma}\;.
$$
The potential is
\be
\lpr = -\fracds{1}{2}\; Q^{+\alpha\beta\gamma}\;Q^+_{\ \alpha\beta\rho} \;
Q^{+\sigma\tau\rho}\;Q^+_{\ \sigma\tau\gamma}\;.
\ee

\vspace{2 mm}
\noindent
4. $F_{4(+4)}\ / Sp(3)\times Sp(1)$. The coset parameters transform as
components of
a traceless antisymmetric tensor of rank 3 for $Sp(3)$. In other
words,
$$
Q^{+\alpha\beta\gamma} =
Q^{+[\alpha\beta\gamma]} \ \ {\rm with}\ \
\Omega_{\alpha\beta}
Q^{+\alpha\beta\gamma} =0\;,
$$
where greek indices refer to the fundamental three-dimensional
representation of $Sp(3)$
and
$\Omega_{\alpha\beta}$
is the defining symplectic form.
Pseudoreality:
$$
*Q^+_{\ \alpha\beta\gamma} = Q^{+\alpha\beta\gamma}\;.
$$
Potential:
\be
\lpr = -\fracds{3}{2}\; Q^{+\alpha\beta\gamma}\;Q^+_{\alpha\beta\rho}\;
Q^{+\sigma\tau\rho}\;Q^+_{\sigma\tau\gamma}\;.
\ee

\vspace{2 mm}
\noindent
5. $E_{6(+2)} \ /SU(6)\times Sp(1)$. In this case the coset parameters
$Q^{+\alpha\beta\gamma}$
are described by the rank 3 antisymmetric tensor of $SU(6)$.
The pseudoreality conditions and the potential
look exactly as in the previous case:
$$
*Q^+_{\ \alpha\beta\gamma} = Q^{+\alpha\beta\gamma}
$$
\be
\lpr = -\fracds{3}{2}\; Q^{+\alpha\beta\gamma}\;Q^+_{\ \alpha\beta\rho}\;
Q^{+\sigma\tau\rho}\;Q^+_{\ \sigma\tau\gamma}\;.
\ee

\vspace{2 mm}
\noindent
6. $E_{7(-5)} \ /SO(12)\times Sp(1)$. The coset parameters
$Q^+_{\ \alpha}$ transform as components of
a chiral spinor of the group $SO(12)$.
Pseudoreality:
$$
*Q^+_{\ \alpha} = Q^{+\alpha}\;.
$$
The potential is
\be
\lpr = \fracds{9}{432}\;
\Sigma_{mn}^{\ \ \alpha\beta}\;
\Sigma_{mn}^{\ \ \gamma\rho}\;
Q^+_{\ \alpha} \;Q^+_{\ \beta} \;Q^+_{\ \gamma} \;Q^+_{\ \rho}\;.
\ee
Here
$$
\Sigma_{mn}^{\ \ \alpha\beta}=
\fracds{1}{2}\left(
\Sigma_{m\ \dot{\beta}}^{\ \alpha}\;
\Sigma_{n}^{\ \dot{\beta}\beta}
- \Sigma_{n\ \dot{\beta}}^{\ \alpha}\;
\Sigma_{m}^{\ \dot{\beta}\beta} \right)\;,
$$
and
$\Sigma_{m\alpha\dot{\beta}}$ are $\sigma$-matrices in 12 dimensions.
Indices are raised and lowered with the charge conjugation
matrix $C^{\alpha\beta}$ which is antisymmetric in 12
dimensions.

\vspace{2 mm}
\noindent
7. $E_{8(-24)} \ / E_7 \times Sp(1)$. The coset parameters
$Q^+_{\ \Sigma}$ are transformed according to the representation
$\underline{56}$ of $E_7$. It is convenient to decompose it
with respect to the maximal subgroup $SU(8) \subset E_7$.
It breaks into antisymmetric tensors
$Q^+_{\ ij}$ and $Q^{+ij}$, where $i$ is the index of the
fundamental representation of $SU(8)$.
Pseudoreality:
$$
*Q^+_{\ ij} = \bar{Q}^{+ij} \;,\;\;\;\; *\bar{Q}^{+ij} = - Q^+_{\ ij} \;.
$$
The potential
takes the form
\be
\lpr = - 4\; Q^+_{\ ik}\; Q^{+il}\; Q^+_{\ nl}\; Q^{+nk} +
        (Q^+_{\ ij}\; Q^{+ij})^2 -
\fracds{1}{4!}\; (\Phi^{+4} + \bar{\Phi}^{+4})\;,
\ee
where
\be
\Phi^{+4} = \epsilon^{i_1 j_1 i_2 j_2 i_3 j_3 i_4 j_4}\;
Q^+_{\ i_1 j_1}\;
Q^+_{\ i_2 j_2}\;
Q^+_{\ i_3 j_3}\;
Q^+_{\ i_4 j_4}\;.
\ee

The general formula for all these $\lpr$s in terms of the structure constants
of $G$ was given in \cite{mal}. It reads as follows
$$
\lpr = \frac{1}{3} (n+2) g^{ab} K^{++}_a K^{++}_b \;.
$$
Here $g^{ab}$ is the restriction of the inverse Killing-Cartan metric of $G$
to $H$ and $K^{++}_a$ is the Killing potential associated with $H$.
The specifity of the cases presented above is encoded (through the structure
constants of $G$) in these objects.

An interesting (though straightforward) problem is to explicitly
compute the quaternionic $\tau$ world metrics for the above $\lpr$'s like this
has been done for the simplest case $Sp(n+1)/Sp(n) \times Sp(1)$ in Sect. 8.

\setcounter{equation}{0}
\section{Conclusions}

In the series of papers including \cite{{a1},{a2}} and the present
article
we have given a complete description of self-dual Yang-Mills theory,
hyper-K\"ahler and quaternionic geometries in the universal language of
harmonic analyticity. The latter underlies both these purely bosonic
theories and their supersymmetric counterparts: $N=2$ Yang-Mills
theory and $N=2$ supergravity \cite{{a3},{a10},{a11}}, thus establishing
a deep affinity between these two classes of theories, so different at first
sight. Moreover, as we have seen in \cite{a2} and in this paper, the
uniform harmonic (super)space description of these theories allows
to understand in a most geometric fashion why
$N=2$ supersymmetry requires the target manifolds of $N=2$ supersymmetric
sigma models to be hyper-K\"ahler or quaternionic. It seems that this
universality  of the harmonic (super)space approach is its main merit.
This approach is most efficient just while exploring the relationships
between the target space geometries and space-time extended supersymmetries.
As an example of its such recent
application we mention the analysis of the target space
geometry of $2D$ $N=(4, 0)$ sigma models in \cite{a17}. In contrast to the
cases considered here and in \cite{a2}, this geometry is non-Riemannian, the
torsion essentially enters into game. However, the harmonic space techniques
proved to be adequate for this case too.

One of intriguing problems for the future study is to find out
possible implications of $N=3$ harmonic analyticity \cite{a4} in the
purely bosonic gauge theories. The geometry underlied by this type of
analyticity (if exists) is expected to be intimately related to
$N=3$ supersymmetric theories: $N=3$ Yang-Mills theory \cite{a4} and
supergravity.

\def\thesection { }
\section{Acknowledgements}
We sincerely thank V.I. Ogievetsky  for interest in the work and
stimulating and useful discussions. A.G. acknowledges
support by the U.S. National Science Foundation, grant PHY-90096198.
E.I. thanks ENSLAPP in Lyon for hospitality.

\end{document}